\documentclass[twocolumn,prstab,superscriptaddress,showpacs,amsmath,amssymb]{revtex4-1}

\usepackage{graphicx}

\usepackage{amssymb}
\usepackage{amsmath}

\usepackage{multirow}

\usepackage{url}

\begin{document}

\title{Target Studies for Surface Muon Production}

\def\PSI{Paul Scherrer Institute (PSI), CHÐ5232 Villigen PSI, Switzerland}
\def\ETH{Institute for Particle Physics, ETH Zurich, 8093 Zurich, Switzerland}

\author{F.~Berg}\affiliation{\PSI}\affiliation{\ETH}
\author{L.~Desorgher}\altaffiliation{Present address: Institut de radiophysique, CHUV, Rue du Grand-Pr\'e 1, CH-1007 Lausanne, Switzerland}\affiliation{\PSI}
\author{A.~Fuchs}\affiliation{\PSI}
\author{W.~Hajdas}\affiliation{\PSI}
\author{Z.~Hodge}\affiliation{\PSI}\affiliation{\ETH}
\author{P.-R.~Kettle}  \email[]{peter-raymond.kettle@psi.ch}\affiliation{\PSI}
\author{A.~Knecht} \email[]{a.knecht@psi.ch}\affiliation{\PSI}
\author{R.~L\"uscher}\affiliation{\PSI}
\author{A.~Papa}\affiliation{\PSI}
\author{G.~Rutar}\affiliation{\PSI}\affiliation{\ETH}
\author{M.~Wohlmuther}\affiliation{\PSI}

\date{\today}

\begin{abstract}
Meson factories are powerful drivers of diverse physics programmes. With beam powers already in the MW-regime attention has to be turned to target and beam line design to further significantly increase surface muon rates available for experiments. 

For this reason we have explored the possibility of using a neutron spallation target as a source of surface muons by performing detailed \textsc{Geant4} simulations with pion production cross sections based on a parametrization of existing data. While the spallation target outperforms standard targets in the backward direction by more than a factor 7 it is not more efficient than standard targets viewed under 90$^\circ$.

Not surprisingly, the geometry of the target plays a large role in the generation of surface muons. Through careful optimization, a gain in surface muon rate of between 30 - 60\% over the standard ``box-like'' target used at the Paul Scherrer Institute could be achieved by employing a rotated slab target. An additional 10\% gain could also be possible by utilizing novel target materials such as, e.g., boron carbide.

\end{abstract}

\pacs{29.25.Rm,	
14.60.Ef	
\vspace{0.5cm}
}


\maketitle

\section{Introduction}
The development of powerful proton drivers in the 1970s enabled a broad experimental programme centred around the various secondary particles produced at dedicated target stations. Proton drivers with energies between 500~MeV and 3000~MeV and currents ranging up to 2.4~mA can be considered as true ``meson factories'' producing up to several $10^8$~$\mu^+$/s or $10^{10}$~$\pi^+$/s through proton nucleus interactions \cite{mesFact}. The experimental programmes range from particle, nuclear and atomic physics experiments with pions and muons \cite{Eri91} to material science experiments using the $\mu$SR technique \cite{Blu99}. While the use of pions has declined over the last decade high-intensity beams of muons are growing in demand. 

Typical experiments make use of the beneficial properties of so-called surface muons \cite{Pif76}. These are copiously produced low-energy muons that can be stopped in extremely thin targets ($\sim$160~mg/cm$^2$). They originate from stopped positive pion decay close to the surface of the production target. By tuning a beam line close to the 2-body decay momentum of 29.8~MeV/c (kinetic energy of 4.1~MeV) muons are selected that escape the target with a momentum $p$ ranging from 0 - 29.8~MeV/c, corresponding to a maximum depth of less than 1~mm in graphite and following a $p^{3.5}$ power law \cite{Pif76}.

Muons above this momentum are also present in the beam but are suppressed by typically two orders of magnitude in this momentum region. These ``cloud muons'' originate from pion decay-in-flight in and around the production target and can have both charge-signs, unlike surface muons which are only positively charged. Hence the limited range of muons from stopped pion decay effectively leads to a very bright and quasi-monochromatic bunched source of surface muons.

While the experiments hunger after even more muons -- especially the searches for lepton flavour violating muon decays (see, e.g., \cite{Ber13}) or the generation of ultra-low energy muons for $\mu$SR applications \cite{Bak04} -- the development of next generation proton drivers with beam powers in excess of the current limit of 1.4~MW still requires significant R\&D. So the attention has turned to the optimization of existing target stations and beam lines and the exploration of novel target ideas. Similar efforts have also been started by another group looking to optimize the target for the ISIS facility at the Rutherford Appleton Laboratory in the United Kingdom \cite{Bun13,Bun14}. With the combined capture and transport efficiencies of traditional beam lines being of $\mathcal{O}(1\%)$ there is certainly a large potential for improvement and first beam lines with larger acceptances have been constructed \cite{Pro08,Str13}. However, this is beyond the scope of this work and we focus here solely on the target side.

The paper is organized as follows: In Section~\ref{sec_XS} we describe the parametrized pion production cross sections that we employ in our simulations for the generation of surface muons. This is followed by Section~\ref{sec_spallationtarget}, in which we explore the possibility of extracting surface muons from an existing spallation target. Section \ref{sec_TgE} then describes an existing standard target for surface muon production followed by the two Sections~\ref{sec_TgEOptimization} and \ref{sec_MaterialOptimization} where we explore the possibilities of enhancing the surface muon production by optimizing the shape and material of the standard target.

\section{Pion Production Cross Sections}\label{sec_XS}
Pions are produced at a proton accelerator through a multitude of different channels. Above the single pion production threshold of $\sim$280~MeV in the centre-of-mass frame the following reactions are available:
\begin{equation*}
\begin{tabular}{lcl}
p + p $\rightarrow$ p + n + $\pi^+$ & & p + n $\rightarrow$ p + n + $\pi^0$ \\
p + p $\rightarrow$ p + p + $\pi^0$ & & p + n $\rightarrow$ p + p + $\pi^-$ \\
p + p $\rightarrow$ d + $\pi^+$ & & p + n $\rightarrow$ n + n + $\pi^+$  \\
& & p + n $\rightarrow$ d + $\pi^0$ 
\end{tabular}
\end{equation*}
Beyond a proton energy of 600~MeV the creation of pairs of pions becomes possible and additional reaction channels open up:
\begin{equation*}
\begin{tabular}{lcl}
p + p $\rightarrow$ p + p + $\pi^+$ + $\pi^-$ & & p + n $\rightarrow$ p + n + $\pi^+$ + $\pi^-$ \\
p + p $\rightarrow$ p + p + $\pi^0$ + $\pi^0$ & & p + n $\rightarrow$ p + n + $\pi^0$ + $\pi^0$ \\
p + p $\rightarrow$ n + n + $\pi^+$ + $\pi^+$ & & p + n $\rightarrow$ n + n + $\pi^+$ + $\pi^0$ \\
p + p $\rightarrow$ n + p + $\pi^+$ + $\pi^0$ & & p + n $\rightarrow$ d + $\pi^-$ + $\pi^+$ \\
p + p $\rightarrow$ d + $\pi^+$ + $\pi^0$ & & p + n $\rightarrow$ d + $\pi^0$ + $\pi^0$ \\
& & p + n $\rightarrow$ p + p + $\pi^-$ + $\pi^0$ \\
\end{tabular}
\end{equation*}
At even higher proton energies further higher multiplicity pion production channels become possible. However, for traditional meson factories with energies below 1000~MeV only the above reaction channels are relevant.

In the early years of the meson factories detailed measurements of the pion production cross sections were performed at SIN (now Paul Scherrer Insitute PSI) and at the 184'' cyclotron of the Lawrence Berkeley National Laboratory LBNL at proton energies of 585~MeV and 730~MeV, respectively \cite{Coc72, Cra80, Cra80b}. Especially the measurements at low pion energies \cite{Cra80b} are of utmost importance in understanding the generation of surface muons.

\begin{figure}[!h]
\centering
\begin{tabular}{c}
  \includegraphics[width=1.0\columnwidth]{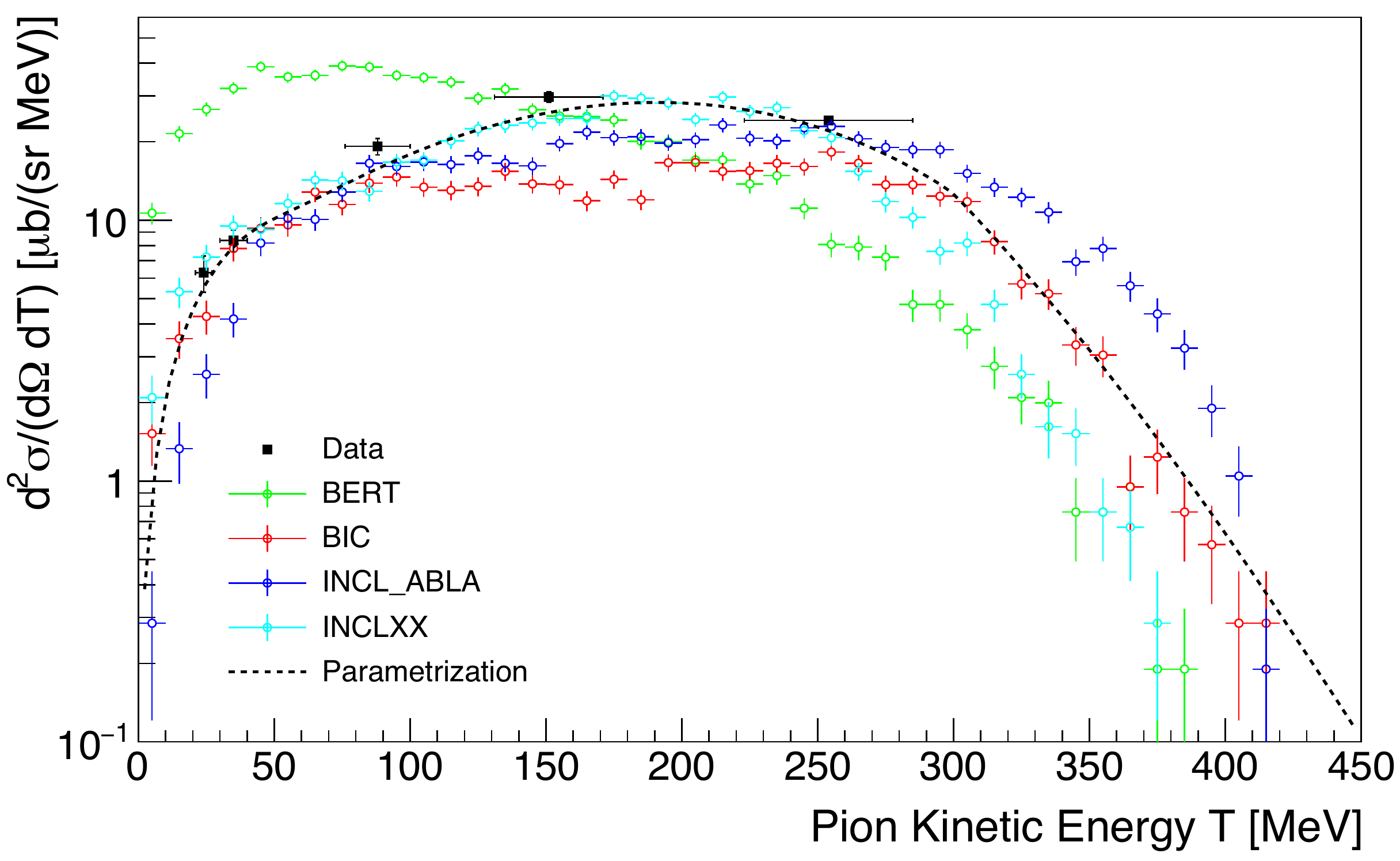} \\
  (a) \\
  \\
   \includegraphics[width=1.0\columnwidth]{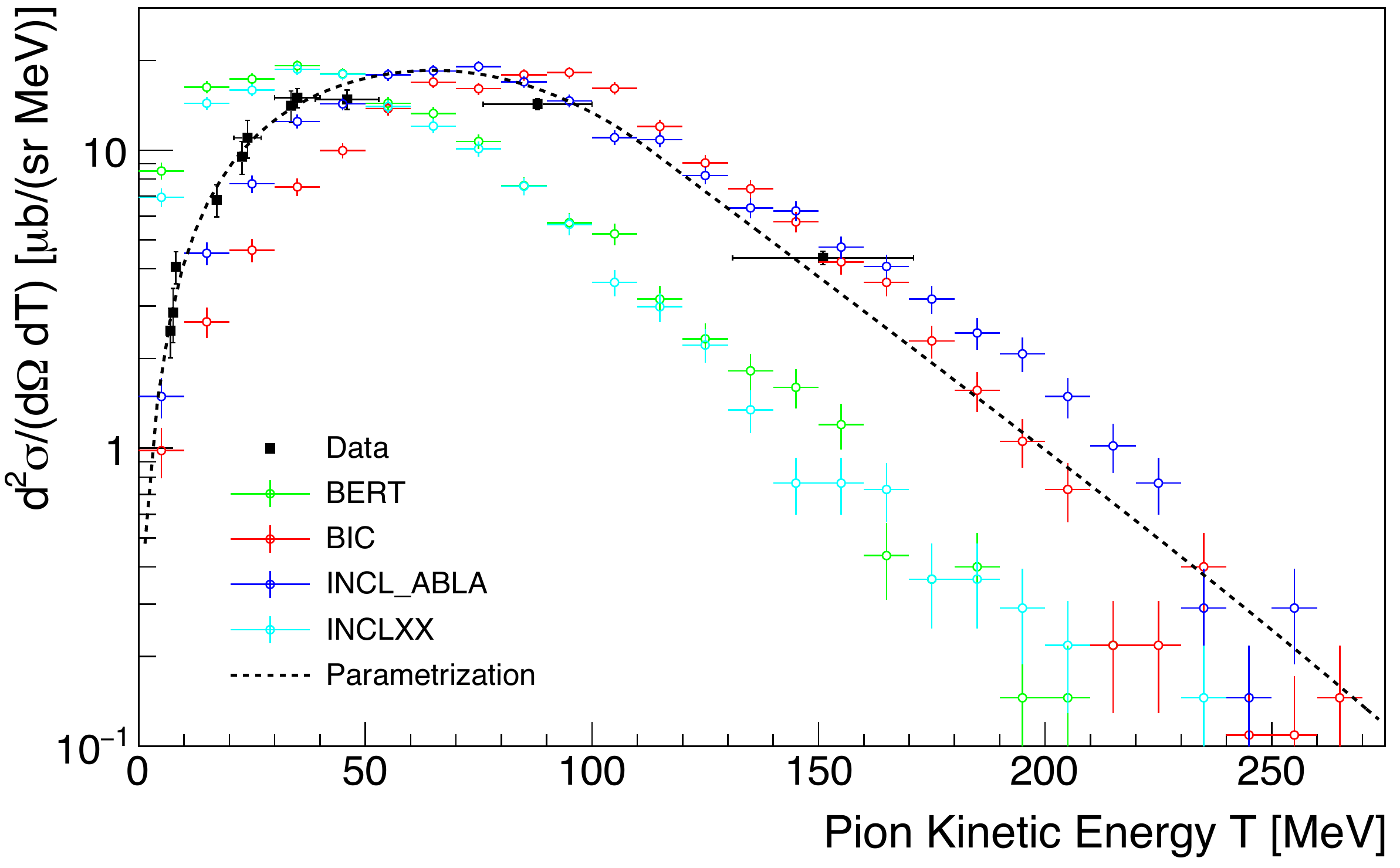} \\
  (b) \\
  \\
   \includegraphics[width=1.0\columnwidth]{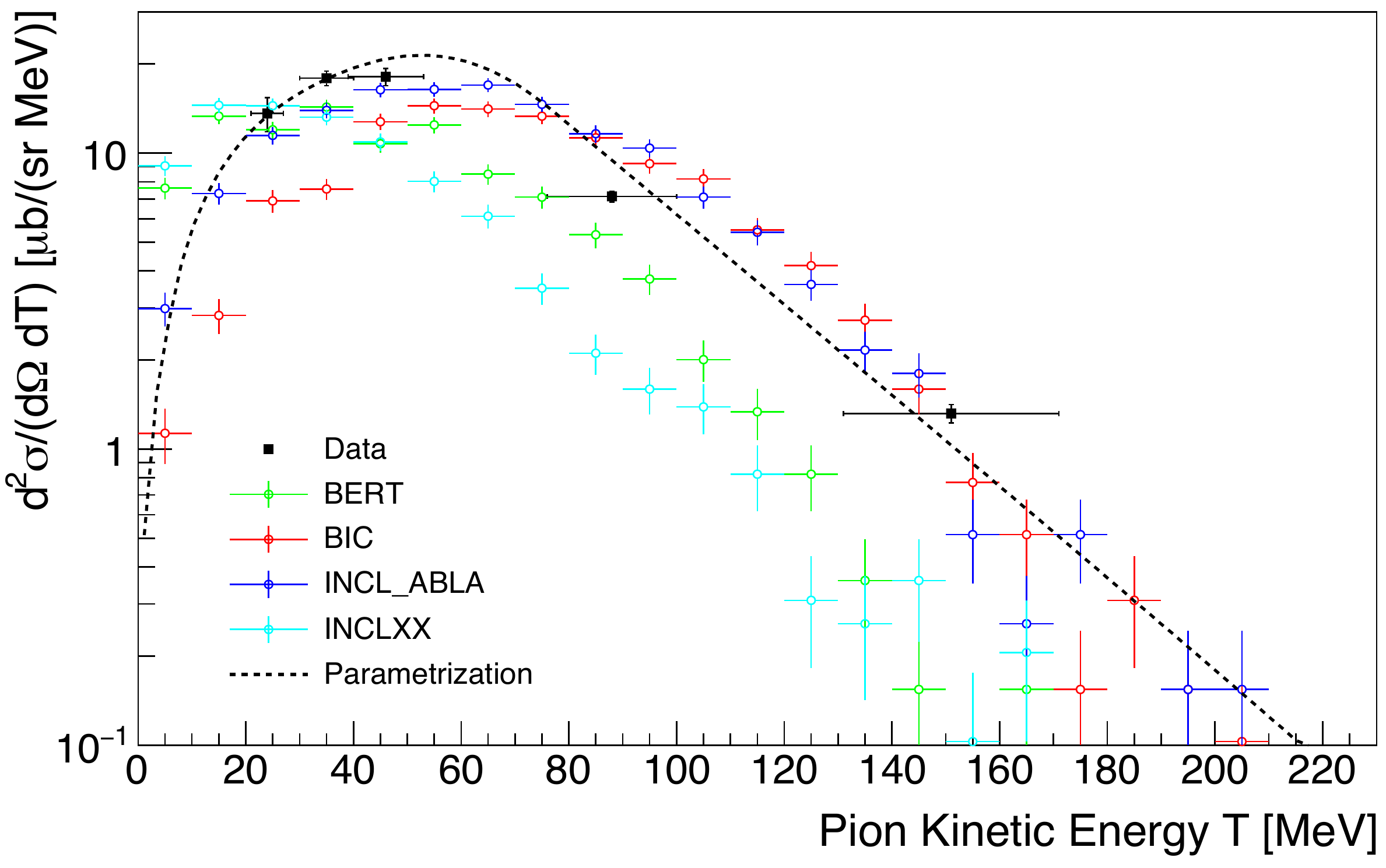} \\
  (c) 
\end{tabular}
\caption{(Color online) Simulated double-differential cross sections for $\pi^+$ production on carbon at a proton energy of 585~MeV and a scattering angle of (a) 22.5, (b) 90 and (c) 135 degrees for several hadronic models used in \textsc{Geant4} 4.9.6 (BERT, BIC, INCLXX) and 4.9.5 (INCL\_ABLA) in comparison to data from \cite{Cra80,Cra80b}. The parametrization is described in the text.}
\label{fig_XSComparison}      
\end{figure}

Hadronic models distributed with \textsc{Geant4} \cite{Geant4} are generally able to model the pion production reactions given above. However, several models perform rather poorly and even models that perform well for certain proton energies, scattering angles and for certain elements perform poorly under other conditions. Figure~\ref{fig_XSComparison} shows a comparison of data with the results of various hadronic models widely used with \textsc{Geant4}. Especially the two models BERT (the default \textsc{Geant4} hadronic model) and INCLXX deviate strongly by as much as a factor of 10 \footnote{A recent study comparing the results of various \textsc{Geant4} versions with measured pion production cross-sections found generally better agreement with older versions of \textsc{Geant4} \cite{Bun14}.}.

For the above reasons we have embarked on the task of introducing reliable $\pi^+$ production cross sections into our \textsc{Geant4} simulations. The basis for our own cross sections rely on two parametrizations found in literature \cite{Bur89,Fro92}. For completeness we present the relevant formulae.

The first parametrization \cite{Fro92} is only valid for low pion kinetic energies $T_{\pi^+}$$\lesssim$40~MeV and reactions on carbon at proton energies of 580~MeV. The double-differential cross section is given by
\begin{equation}\label{eq_XSle}
\frac{\mathrm{d}^2\sigma_\mathrm{LE}}{\mathrm{d}\Omega \mathrm{d}T_{\pi^+}} = S_{01} \sin \left( \frac{\pi T_{\pi^+}}{2T_{01}} \right) - S_{02} \sin \left( \frac{\pi T_{\pi^+}}{2T_{02}} \right) \cos \theta
\end{equation}
with the parameters $S_{01} = 15.3$~$\mu$b/(sr MeV), $S_{02} = 5.6$~$\mu$b/(sr MeV), $T_{01} = 49.4$~MeV,  $T_{02} = 32.4$~MeV, and the angle $\theta$ being between the momenta of proton and produced $\pi^+$.

The second parametrization \cite{Bur89} is valid for all elements, proton energies $T_p < 800$~MeV and all pion energies. However, as shown later it will only be used at pion energies above approximately 40~MeV as it performs rather poorly below that energy. The basic shape is modeled by a Gaussian function with a high-energy cut-off and parameters fitted to data. In addition, an amplitude based on B-splines \cite{B-spline} controls the overall normalization.
\begin{eqnarray}\label{eq_XShe}
\frac{\mathrm{d}^2\sigma_\mathrm{HE}}{\mathrm{d}\Omega \mathrm{d}T_{\pi^+}} & = & A(\theta,Z,T_p) \exp \left[- \left( \frac{\bar{T}(\theta,Z,T_p) - T_{\pi^+}}{\sqrt{2} \sigma (\theta,Z,T_p)}  \right)^2 \right]   \nonumber \\
& &  \frac{1}{1+ \exp \left[ \frac{T_{\pi^+} - T_F}{B} \right]}
\end{eqnarray}
The exact details of the different parameters and a small change in the basic shape of the parametrization can be found in the Appendix.

In order to get a good description over the full range of pion energies the two parametrizations are combined using a smooth sigmoid function in order to move from the low-energy to the high-energy regime. Additionally, the low-energy parametrization of Eq.~(\ref{eq_XSle}) is scaled for arbitrary elements and proton energy using the high-energy parametrization. The combined differential cross section is then given by
\begin{eqnarray}\label{eq_XS}
\frac{\mathrm{d}^2\sigma}{\mathrm{d}\Omega \mathrm{d}T_{\pi^+}} (T_{\pi^+}, \theta) & = & (1-f_{t}) f_{s} \frac{\mathrm{d}^2\sigma_\mathrm{LE}}{\mathrm{d}\Omega \mathrm{d}T_{\pi^+}} (T_{\pi^+} - T_{\pi^+}^0 , \theta) \nonumber \\
& & + f_{t} \frac{\mathrm{d}^2\sigma_\mathrm{HE}}{\mathrm{d}\Omega \mathrm{d}T_{\pi^+}} (T_{\pi^+}, \theta)
\end{eqnarray}
with the pion kinetic energy $T_{\pi^+}$ in MeV, the sigmoidal transition function given by
\begin{equation}
f_t = \frac{1}{1 + \exp \left( - (T_{\pi^+} - 40)/10 \right)}
\end{equation}
and the scaling factor by 
\begin{eqnarray}
f_s & = & \left. \frac{\mathrm{d}^2\sigma_\mathrm{HE}}{\mathrm{d}\Omega \mathrm{d}T_{\pi^+}} (T_{\pi^+}=40, \theta=90^\circ, Z, T_p) \right/  \nonumber \\
& & \frac{\mathrm{d}^2\sigma_\mathrm{HE}}{\mathrm{d}\Omega \mathrm{d}T_{\pi^+}} (T_{\pi^+}=40, \theta=90^\circ, Z=6, T_p=585) \,.
\end{eqnarray}
The shift in pion kinetic energy by $T_{\pi^+}^0$ was observed in Ref.~\cite{Cra80b} and is attributed to the coulomb repulsion between the nucleus and the pion. Essentially the coulomb force of the nucleus imparts a minimal kinetic energy on the pion. The effect for nickel was measured to be about 5~MeV. By scaling the coulomb potential and size of the nucleus the shift is thus given for an arbitrary nucleus of atomic number $Z$ and mass number $A$ \footnote{For practical reasons we use in our simulations the simplification $A \sim 2Z$. \\ As the mathematical prescription given in Eq.~(\ref{eq_XS}) results in negative cross section values for $T_{\pi^+} < T_{\pi^+}^0$ we set them explicitly to zero.} by
\begin{equation}
T_{\pi^+}^0 = 0.696 Z/A^{1/3} \quad \left[ \mathrm{MeV} \right]
\end{equation}
with the numerical constant adjusted to match the shift measured for nickel.

The result of the combined parametrization can be seen in Fig.~\ref{fig_XSComparison}. The parametrization shows generally good agreement over the full parameter space. The accuracy of the parametrization is of the order 10\% -- the accuracy of the measured cross sections.

In addition, we employ Monte-Carlo biasing/splitting techniques in our simulations. At each point of a hadronic interaction that produces a $\pi^+$ we typically create 100 $\pi^+$, each randomly sampled from the relevant distribution. The same technique is employed at the point of $\pi^+$ decay where again typically 100 $\mu^+$ are generated. In the end each $\mu^+$ produced corresponds to $10^4$ protons on target.

\section{Spallation Target}\label{sec_spallationtarget}
\begin{figure}[!h]
\includegraphics[width=0.65\columnwidth]{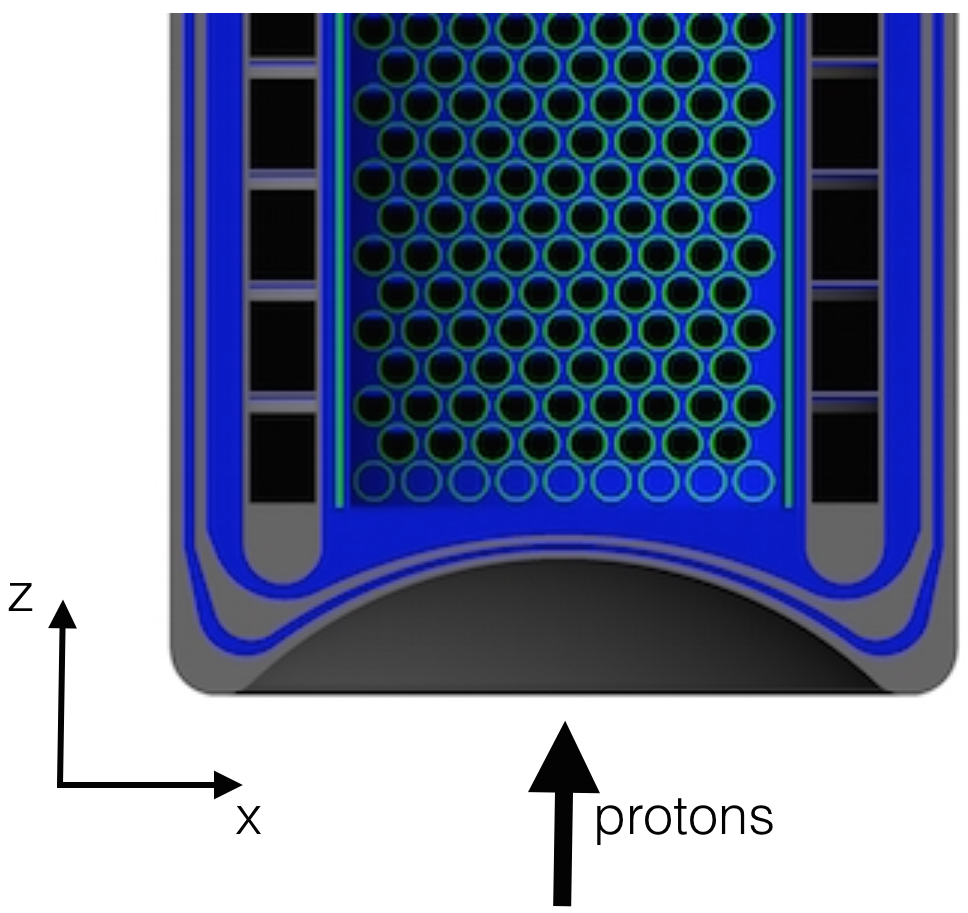}
\caption{(Color online) Cross section of the SINQ spallation target as implemented in our \textsc{Geant4} simulations. Visible are the outer and inner safety vessels made of AlMg$_3$ (grey), the zircaloy (green) rods filled with lead (black) and the lead reflector surrounding the zircaloy rods. Protons enter the target from below.}
\label{fig_SINQtarget}      
\end{figure}

The original motivation for implementing reliable pion production cross sections into \textsc{Geant4} was the idea of using a spallation target as a source for surface muons \cite{Ket10,Blo13}. The basis for our simulations is given by the target of the Swiss Spallation Neutron Source (SINQ) \cite{SINQ} located at the Paul Scherrer Insitute. Figure~\ref{fig_SINQtarget} shows the spallation target as implemented in the simulations. The target is a so-called ``Cannelloni'' target where the lead used for the spallation process is enclosed in individual zircaloy rods. In the case of the SINQ target a lead reflector also surrounds the rods. The target is cooled by heavy water and enclosed in a double-walled AlMg$_3$ safety vessel with a concave window through which the protons impinge on the target. The outer diameter of the safety vessel is 212~mm.

The constraints around a spallation target are typically very severe. Not only in terms of the massive radiation load that any muon capture element would have to withstand but also in terms of space. In the case of SINQ, e.g., the spallation target is inserted into a beam pipe of only 220~mm diameter surrounded by the heavy water moderator. Any muon capture element is therefore required to fit inside this beam pipe.

While Fig.~\ref{fig_SINQtarget} shows the full geometry with each individual lead filled zircaloy rod implemented, usually the simulation was, for reasons of computational speed, performed with the individual zircaloy rods replaced by a simple appropriate mixture of materials. No difference was observed in the results from the two approaches.

The proton beam parameters were taken from fits to the simulation of the proton beam line at PSI performed using TRANSPORT~\cite{TRAN} and TURTLE~\cite{TURT}. The horizontal and vertical beam size ($\sigma_x$ and $\sigma_y$) and proton beam current $I_p$ at the entrance window are
\begin{eqnarray}\label{eq_protonbeam}
\sigma_x & = & 21.4 \, \mathrm{mm} \nonumber \\
\sigma_y & = & 29.6 \, \mathrm{mm} \nonumber \\
I_p & = & 1.7  \, \mathrm{mA} \, .
\end{eqnarray}
The beam divergence was found to have a negligible impact on the results and was set to zero.

\begin{figure}[!h]
\centering
\begin{tabular}{c}
  \includegraphics[width=1.0\columnwidth]{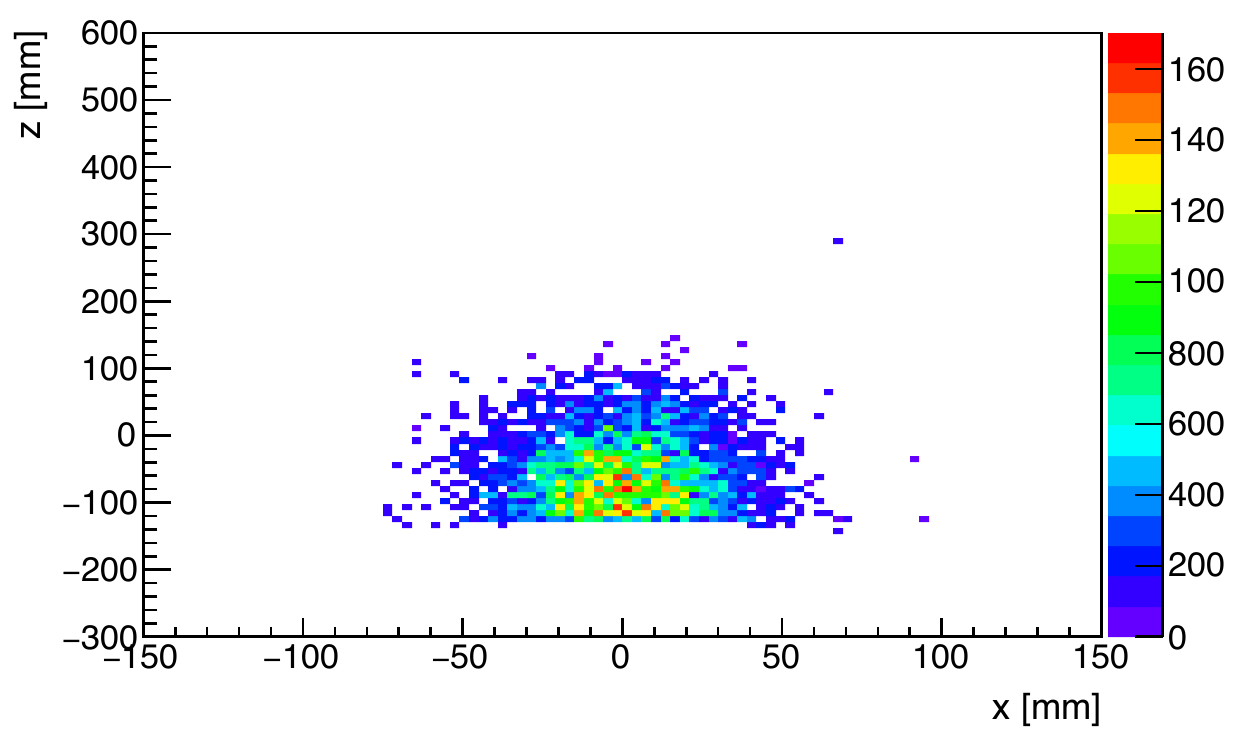} \\
  (a) \\
  \\
   \includegraphics[width=1.0\columnwidth]{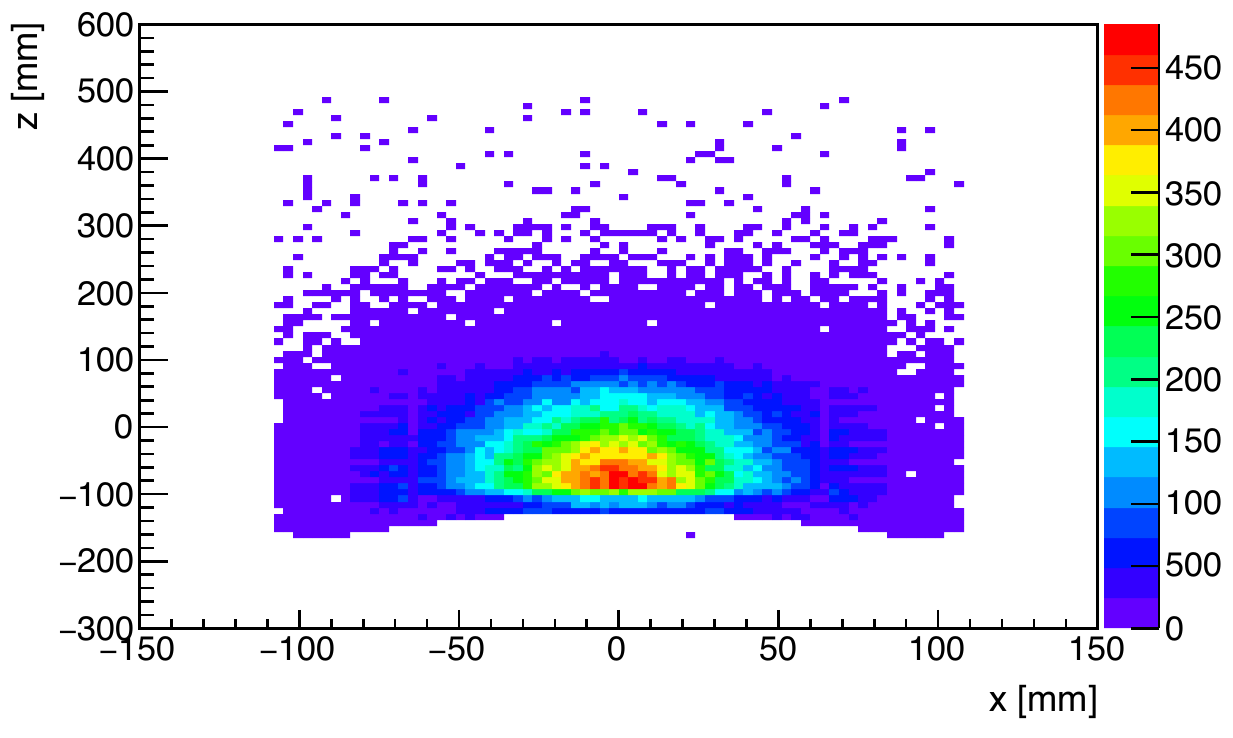} \\
  (b) 
\end{tabular}
\caption{Production vertices for (a) pions and (b) muons from pion decay at rest in a central vertical slice of 10~mm thickness through the spallation target. The lowest position of the safety vessel is at $z\approx -160$~mm with the concave window extending up to $z\approx -125$~mm.}
\label{fig_ProductionVertices}      
\end{figure}

Figure~\ref{fig_ProductionVertices} shows the production vertices of pions and muons from pion decay at rest, from within a 10~mm thick vertical slice through the centre of the spallation target. While the initial pion production positions are still fairly close together and feature a sharp cut-off at around $z \approx 100$~mm due to the proton energy dropping below the pion production threshold, the muon positions are spread out and are governed by the material distribution and the corresponding pion range.

\begin{figure}[!h]
\centering
\begin{tabular}{c}
  \includegraphics[width=1.0\columnwidth]{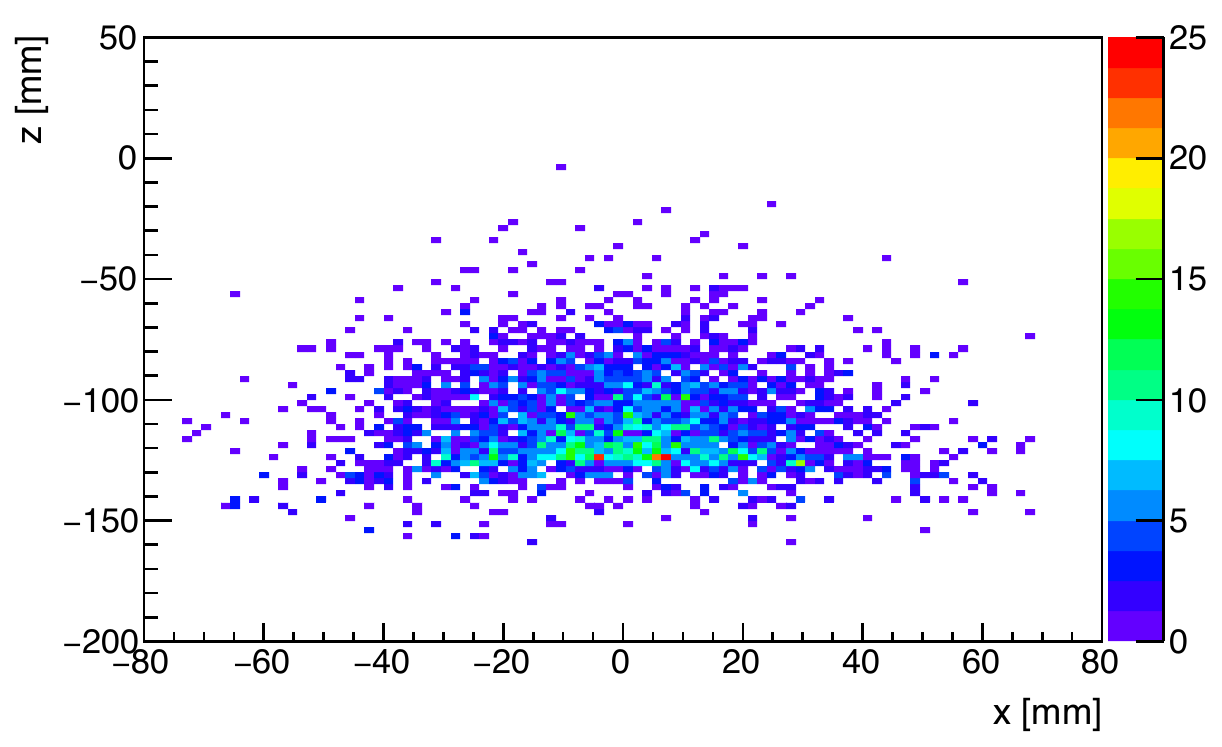} \\
  (a) \\
  \\
   \includegraphics[width=1.0\columnwidth]{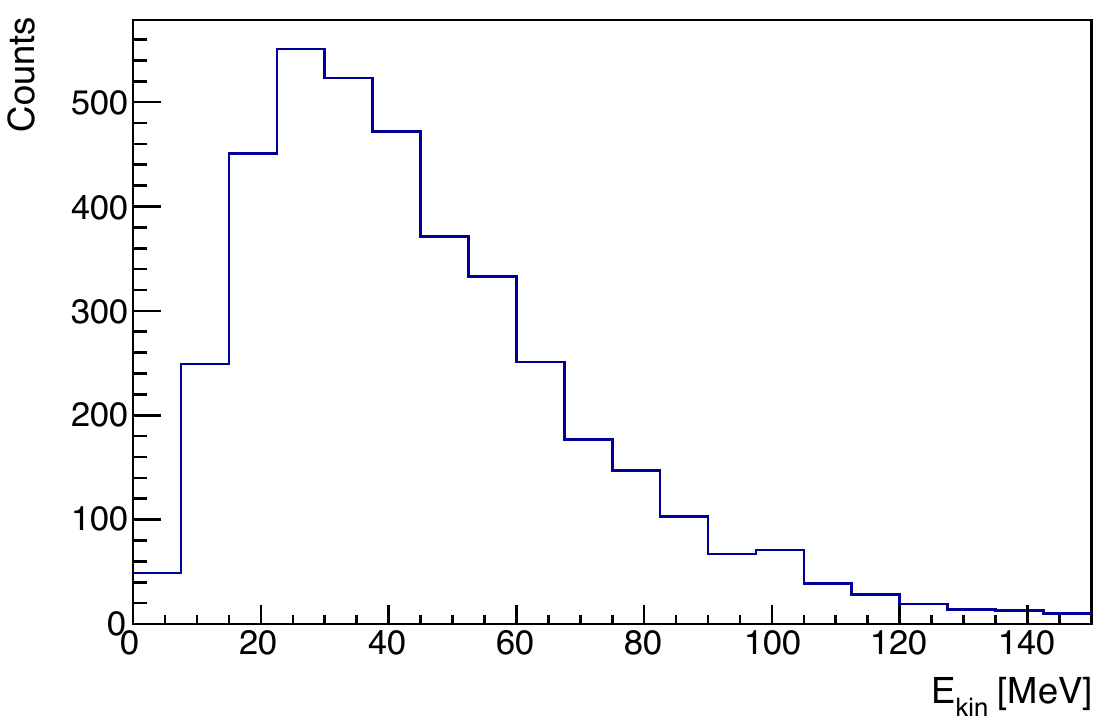} \\
  (b) 
\end{tabular}
\caption{Initial position (a) and kinetic energy (b) of pions that generate detected surface muons. Please note the changed scale of the y-axis in Fig.~\ref{fig_MotherPions}a) compared to Fig.~\ref{fig_ProductionVertices}.}
\label{fig_MotherPions}      
\end{figure}

In order to assess and characterize the particles leaving the spallation target in our simulation we placed a virtual detector with a slightly larger diameter of 220~mm just below the lowest point of the safety vessel. When mentioning detected particle in the following paragraphs it refers to particles sampled by this detector. In addition this allows one to extract the pion properties that lead to a detected muon. Figure~\ref{fig_MotherPions} shows the initial positions of these pions and their kinetic energy. There are two main areas generating the pions that later lead to surface muons. The first area is the window of the safety vessel itself where low energy pions are immediately stopped close to their initial position. The second area is located at around $z\approx -100$~mm where the proton beam starts impinging on the zircaloy rods and a somewhat larger number of pions is produced. Pions from deeper within the spallation target are effectively shielded by the high stopping power of lead and do not reach the window. The mean energy of the pions contributing to the surface muons is $\sim$50~MeV with a large tail to higher energies (see Fig.~\ref{fig_MotherPions}b)).

\begin{figure}[!h]
\centering
\begin{tabular}{c}
  \includegraphics[width=1.0\columnwidth]{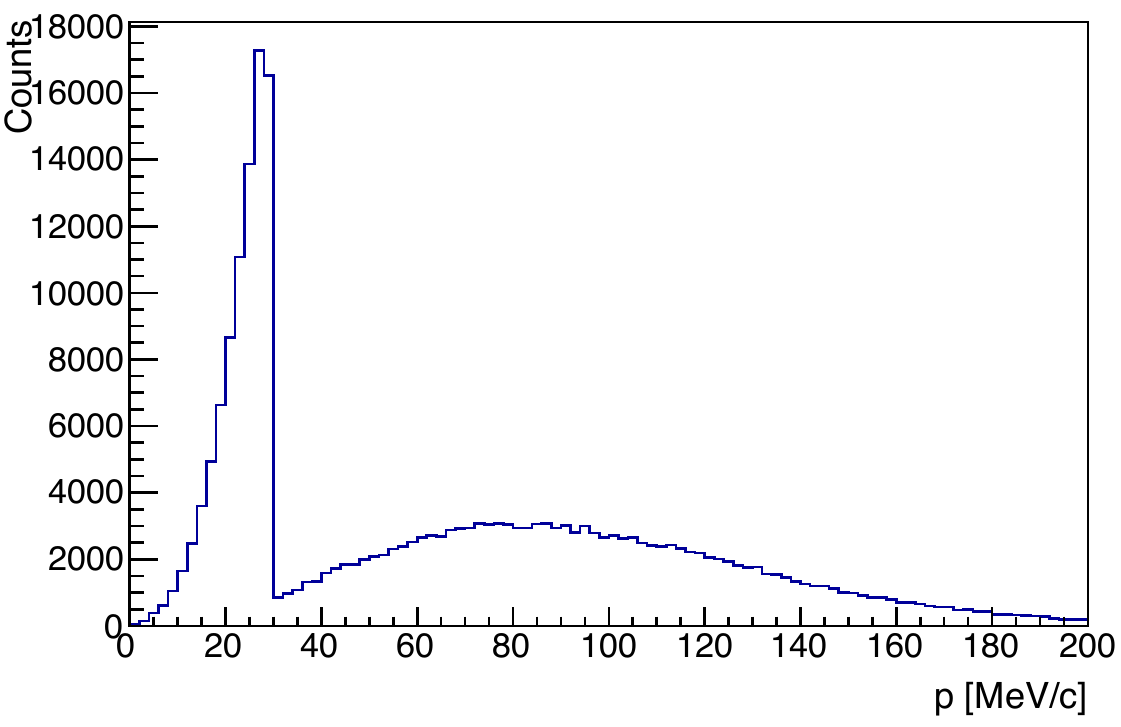} \\
  (a) \\
  \\
   \includegraphics[width=1.0\columnwidth]{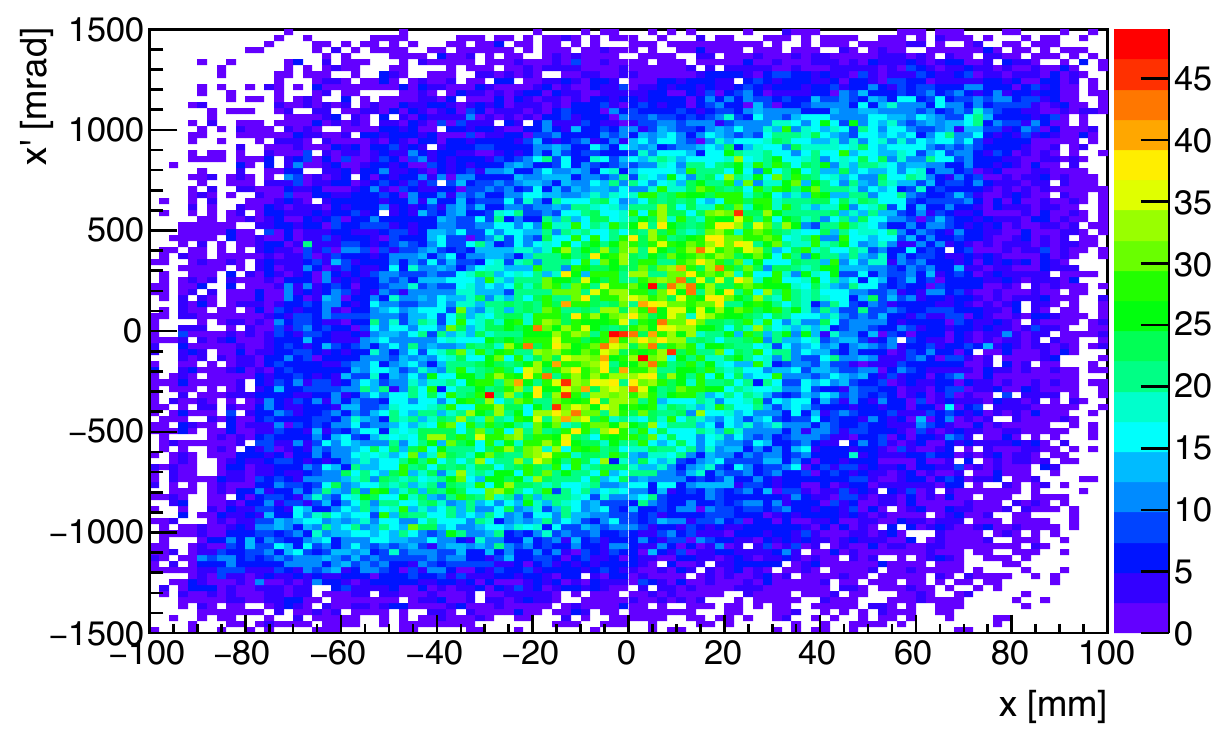} \\
  (b) 
\end{tabular}
\caption{a) Momentum spectrum of all positive muons leaving the spallation target in downward direction. Clearly visible is the peak at 30~MeV/c from surface muons and the broad distribution peaking at around 85~MeV/c from pions decaying in-flight. b) Size and divergence distribution $x$ and $x'$. The $y$-components look similar.}
\label{fig_SourceCharacteristics}      
\end{figure}

Figure~\ref{fig_SourceCharacteristics} shows some characteristics of the spallation target as a source of surface muons. Figure~\ref{fig_SourceCharacteristics}a) gives the momentum spectrum of all muons traversing the virtual detector in downwards direction. Clearly visible are the peaks at $\sim$30~MeV/c stemming from surface muons and the broad peak at $\sim$85~MeV/c coming from muons from pion decay in flight. Figure~\ref{fig_SourceCharacteristics}b) shows the size and divergence of surface muons at the detector along the $x$-axis. The $y$-components look similar. The root-mean-square values are:
\begin{eqnarray}
x_\mathrm{rms} & = & 39.4 \, \mathrm{mm} \nonumber \\
x'_\mathrm{rms} & = & 639 \, \mathrm{mrad} \nonumber \\
y_\mathrm{rms} & = & 42.6 \, \mathrm{mm} \nonumber \\
y'_\mathrm{rms} & = & 642 \, \mathrm{mrad}
\end{eqnarray}
The corresponding phase space is thus large. One of the reasons for the large root-mean-square values in $x$ and $y$ is obviously the large proton beam impinging on the target (see Eq.~(\ref{eq_protonbeam})) \footnote{The proton beam is specifically blown up in front of the spallation target in order to not have a too high energy density at the entrance window.}. This is one of the disadvantages of using a spallation target as a source of surface muons.

Finally, the number of muons per proton with momentum $< 29.8$~MeV/c leaving the spallation target in a downwards direction amounts to
\begin{equation}\label{eq_spallationmp}
\mu^+ / p = 8.8 \times 10^{-6}
\end{equation}
where $8.5 \times 10^{-6}$ stem from surface muons and $ 0.3 \times 10^{-6}$ -- or 3\% -- from pion decay in flight. With the proton beam current of Eq.~(\ref{eq_protonbeam}) the resulting muon rate is 
\begin{equation}
I_{\mu^+} = 9.4 \times10^{10}\, \mu^+/\mathrm{s} \,.
\end{equation}
For a momentum byte of $25.0 < p < 29.8$~MeV/c (typical for high-rate surface muon beams) the rate reduces to
\begin{equation}
I_{\mu^+} = 4.3 \times10^{10}\, \mu^+/\mathrm{s} \,.
\end{equation}

Owing to the fact that the spallation target does not yield higher surface muon rates than a standard target viewed at $90^\circ$ (see Section~\ref{sec_TgE}) together with the large initial phase space and the severe space constraints around the spallation target, leads us to the conclusion that abandoning this original idea and pursuing instead the optimization of standard meson production targets is a better approach.

\section{Standard Meson Production Target: Target E at PSI}\label{sec_TgE}
\begin{figure}
\includegraphics[width=0.5\columnwidth]{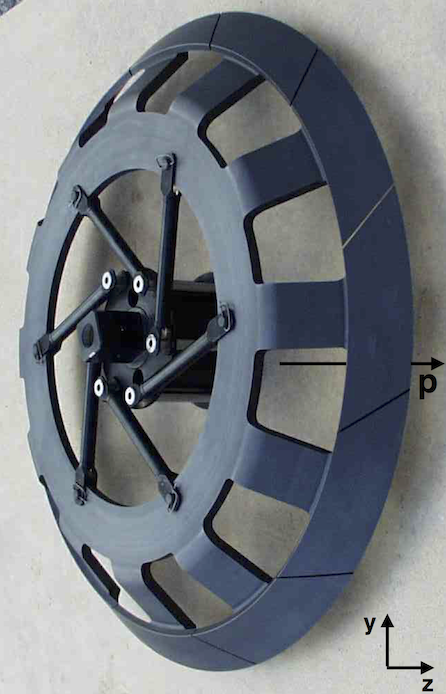}
\caption{Picture of the Target E wheel used for surface muon production at PSI. The proton beam impinges on the outer rim as shown by the arrow. In order to radiatively cool the target, the wheel rotates with a frequency of 1~Hz.}
\label{fig_TgE}      
\end{figure}

In order to make a comparison to the results of the spallation target described above we also simulated a standard meson target, Target E at PSI as shown in Figure~\ref{fig_TgE}. As the proton beam size is small when impinging on the target it is not necessary to simulate the full target wheel and we approximate it by a rectangular box structure of 40~mm length, 6~mm width and 40~mm height. The proton beam impinges on the 6~mm wide face with parameters
\begin{eqnarray}\label{eq_protonbeam_TgE}
\sigma_x & = & 0.75 \, \mathrm{mm} \nonumber \\
\sigma_y & = & 1.25 \, \mathrm{mm} \nonumber \\
I_p & = & 2.4  \, \mathrm{mA} \, .
\end{eqnarray}
As in the case of the spallation target the divergences have a negligible impact and are neglected. The target material is polycrystalline graphite with a density of 1.84~g/cm$^3$.

\begin{figure}
\centering
\begin{tabular}{c}
  \includegraphics[width=1.0\columnwidth]{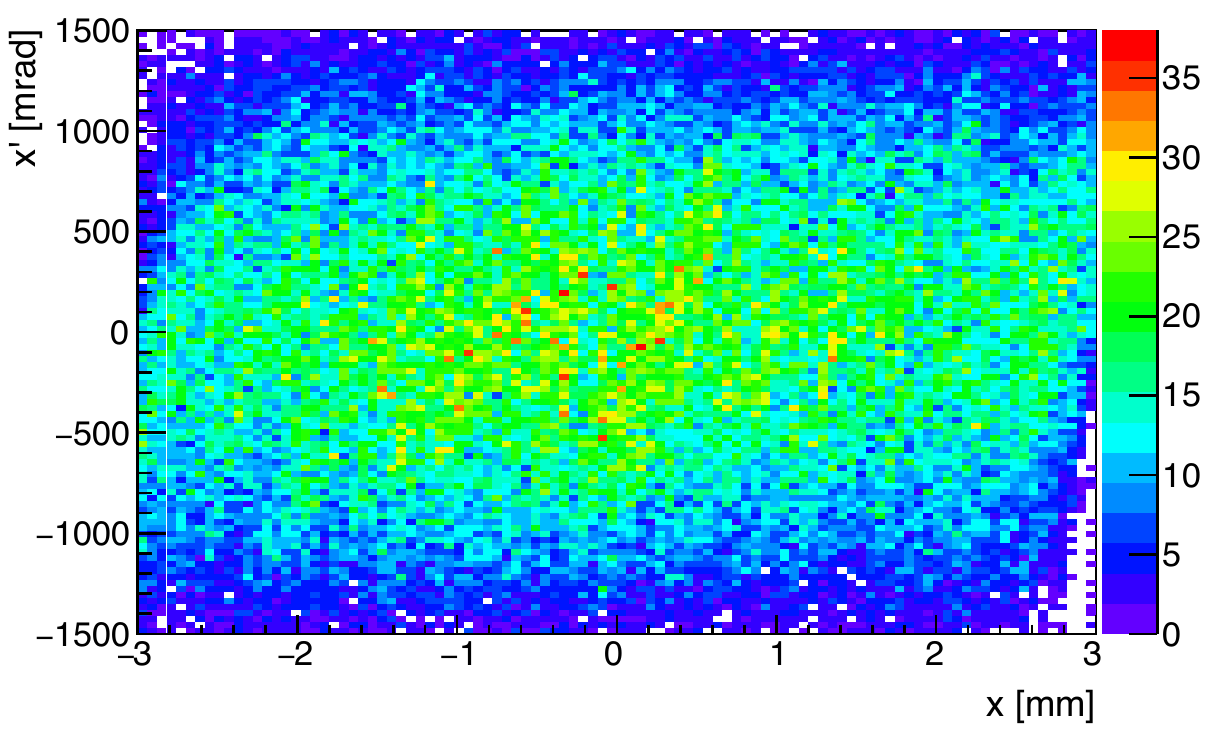} \\
  (a) \\
  \\
   \includegraphics[width=1.0\columnwidth]{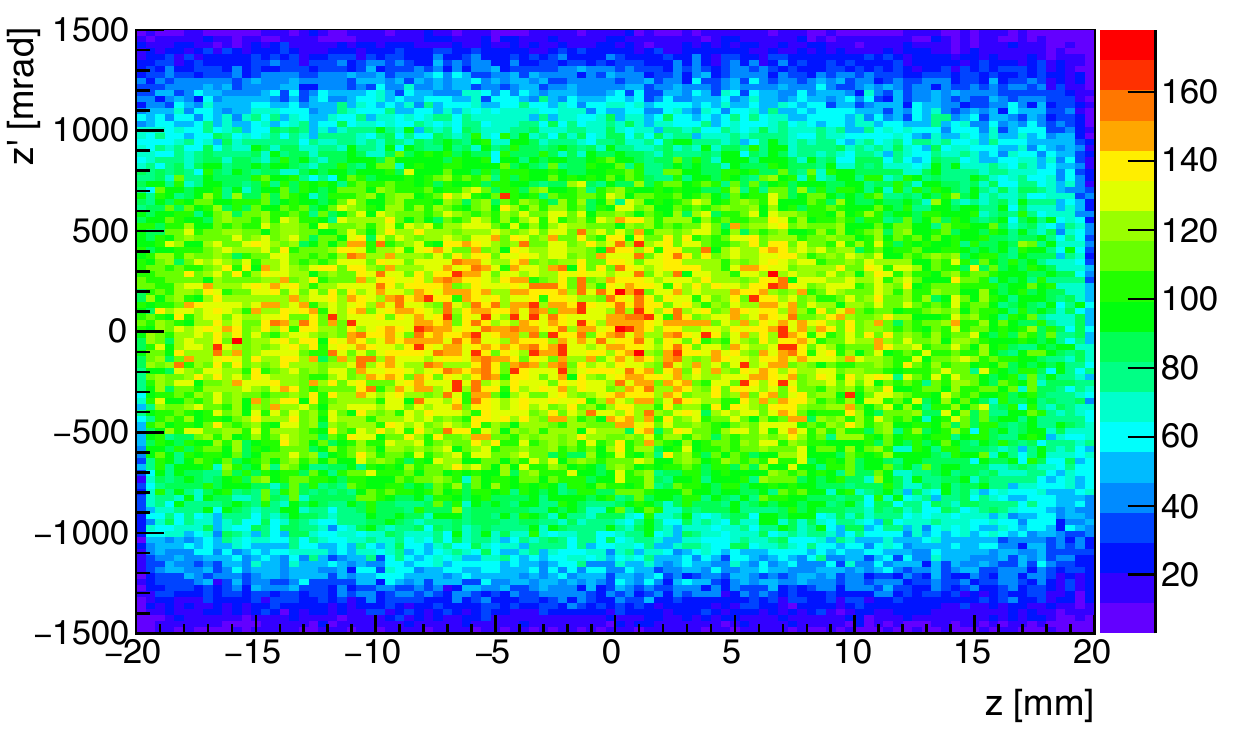} \\
  (b) 
\end{tabular}
\caption{Size and divergence distribution for the backward face (a) and side face (b) of Target E.}
\label{fig_TgECharacteristics}      
\end{figure}

We examined the source characteristics by placing a virtual detector close to each of the four side surfaces and sampling the muons traversing those detectors. Figure~\ref{fig_TgECharacteristics} shows the characteristics of two of the sides. The parameters for the backward face for positive muons with momenta below 29.8~MeV/c are: 
\begin{eqnarray}
x_\mathrm{rms} & = & 1.6 \, \mathrm{mm} \nonumber \\
x'_\mathrm{rms} & = & 668 \, \mathrm{mrad} \nonumber \\
y_\mathrm{rms} & = & 7.3 \, \mathrm{mm} \nonumber \\
y'_\mathrm{rms} & = & 677 \, \mathrm{mrad}  \nonumber \\
\mu^+/p & = & 1.2 \times 10^{-6}  \nonumber \\
I_{\mu^+} & = & 1.8 \times 10^{10}\, \mu^+/\mathrm{s}  
\end{eqnarray}
The values for the forward face are very similar albeit with a reduced muon flux of $1.2 \times 10^{10}$~$\mu^+$/s. For each of the two side faces the values are:
\begin{eqnarray}\label{eq_TgESidemp}
z_\mathrm{rms} & = & 10.9 \, \mathrm{mm} \nonumber \\
z'_\mathrm{rms} & = & 678 \, \mathrm{mrad} \nonumber \\
y_\mathrm{rms} & = & 7.9 \, \mathrm{mm} \nonumber \\
y'_\mathrm{rms} & = & 678 \, \mathrm{mrad}  \nonumber \\
\mu^+/p & = & 8.3 \times 10^{-6}  \nonumber \\
I_{\mu^+} & = & 1.2 \times 10^{11}\, \mu^+/\mathrm{s}  
\end{eqnarray}
Both $z'$ and $y'$ are calculated with respect to the $x$-axis, so $z' = \frac{\mathrm{d}z}{\mathrm{d}x}$ and $y'= \frac{\mathrm{d}y}{\mathrm{d}x}$. It is interesting to note that the side face of Target E is as efficient in generating surface muons as the spallation target (compare the muon to proton numbers of Eq.~(\ref{eq_spallationmp}) and (\ref{eq_TgESidemp})) though featuring a much smaller initial phase space. In the backward direction the spallation target is more than seven times as efficient profiting from its larger size.

\begin{figure}
\includegraphics[width=1.0\columnwidth]{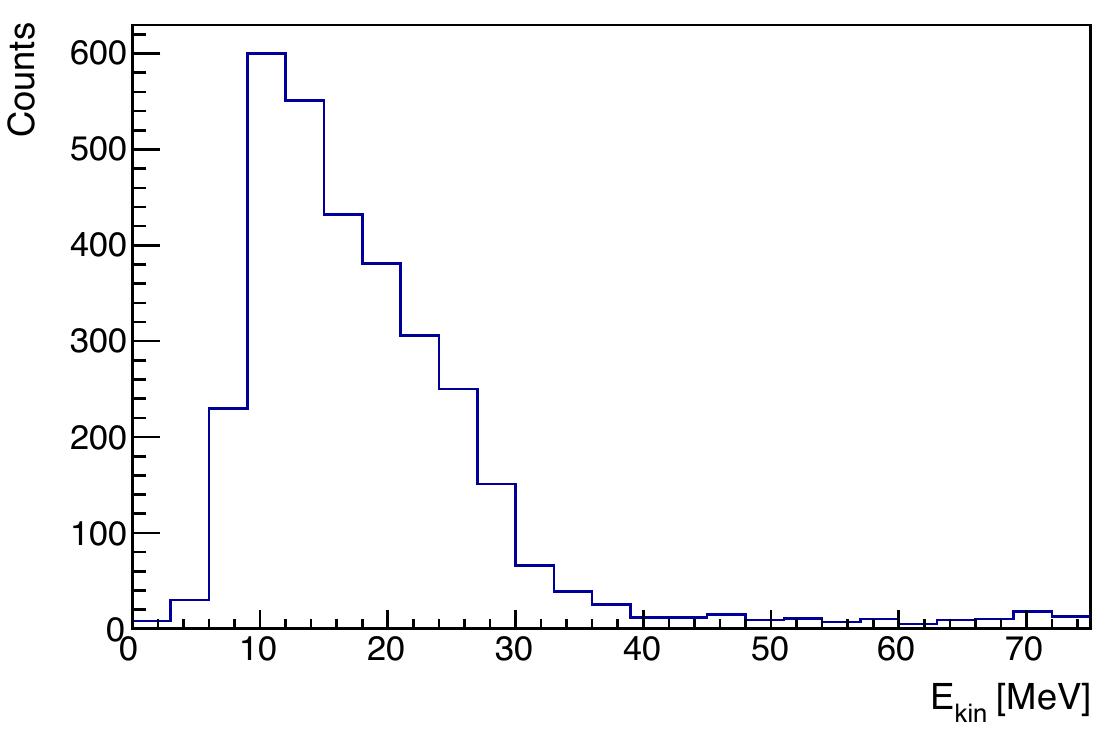}
\caption{Distribution of pion energies that generate surface muons in Target E.}
\label{fig_TgESidePionEnergy}      
\end{figure}

Figure~\ref{fig_TgESidePionEnergy} shows the energy spectrum of pions that generate surface muons at the side face of Target E. The spectrum for the backward and forward faces are similar. The average energy is 18.6~MeV and thus substantially lower than in the case of the spallation target  (see Fig.~\ref{fig_MotherPions}b)).

Table~\ref{tab_TgE_Rates} summarizes the results for the rates obtained by changing the length of Target E but otherwise keeping all parameters the same. While the backward and forward rates saturate above a length of $\sim$20~mm the rates from the side scale more than linearly with the target length. This is due to the fact that the pion stop density within the target needs a certain distance at the beginning and at the end in order to build up to its maximal value (as can be seen in Fig.~\ref{fig_TgECharacteristics}b).

\begin{table}
\centering
\begin{tabular}{c  c c c}
\hline
\hline
Length & Backward & Forward & Side \\
\hline
10 & $1.4\times 10^{10}$ & $9.0\times 10^{9}$ & $1.8\times 10^{10}$ \\
20 & $1.6\times 10^{10}$ & $1.2\times 10^{10}$ & $5.1\times 10^{10}$ \\
30 & $1.9\times 10^{10}$ & $1.1\times 10^{10}$ & $8.5\times 10^{10}$ \\
40 & $1.8\times 10^{10}$ & $1.1\times 10^{10}$ & $1.2\times 10^{11}$ \\
60 & $1.8\times 10^{10}$ & $1.2\times 10^{10}$ & $2.1\times 10^{11}$ \\
\hline
\hline
\end{tabular}
\caption{Surface muon rates in $\mu^+$/s for all muons with momenta below 29.8~MeV/c emitted from the various sides of Target E for various lengths of the target in mm. The values for the side rates correspond to a single side only.}
\label{tab_TgE_Rates}
\end{table}

\section{Optimization of Standard Meson Production Targets}\label{sec_TgEOptimization}
\begin{figure}
\includegraphics[width=1.0\columnwidth]{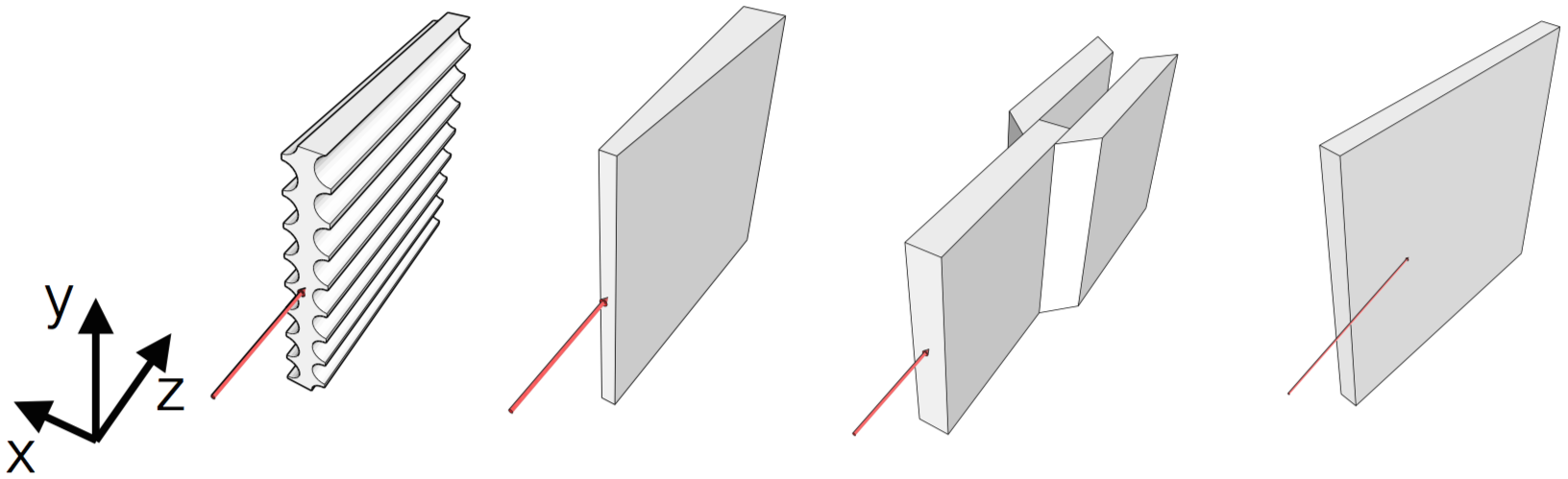}
\caption{(Color online) Different geometries studied in our target optimization. From left to right: grooved target, trapezoidal target, fork target, rotated slab target. The red line marks the proton beam.}
\label{fig_TgEGeometries}      
\end{figure}

Several alternative target geometries were investigated in an attempt to enhance the surface muon production, each with varying degrees of success. These geometries focused on methods of either increasing the surface volume (surface area times acceptance depth) or the pion stop density near the surface. Each geometry was required to preserve, as best as possible, the proton beam characteristics downstream of the target station (spallation neutron source requirement). The muon beam extraction directions considered here are sideways, backwards, and forwards with respect to the proton beam. The accepted phase space used in our simulations roughly corresponds to the acceptance of the following beam lines at PSI: $\mu$E4 (sideways at 90$^\circ$) with a maximum surface muon intensity of $4.8 \times 10^8$~$\mu^+$/s~\cite{Pro08}, $\pi$E5 (backwards at 165$^\circ$) with a maximum surface muon intensity of $1.1 \times 10^8$~$\mu^+$/s~\cite{Ada13} and  $\pi$E1 (forwards at 8$^\circ$) with a maximum surface muon intensity of around $10^6$~$\mu^+$/s~\cite{Wal94} \footnote{The remaining two beam lines around target E at PSI are $\pi$E3 and $\mu$E1. The beam line $\pi$E3 is -- as $\mu$E4 -- located at 90$^\circ$ and will correspondingly see the same enhancement factors. The beam line $\mu$E1 is special as it features a muon decay channel that accepts high energy pions from the target. The rate of high energy pions is only marginally altered by the geometrical changes studied here and $\mu$E1 will thus not be affected.}. All enhancements listed below are relative to the standard target geometry described in Sec.~\ref{sec_TgE}. A model of each geometry investigated is shown in Fig.~\ref{fig_TgEGeometries}.

\begin{figure}
\includegraphics[width=1.0\columnwidth]{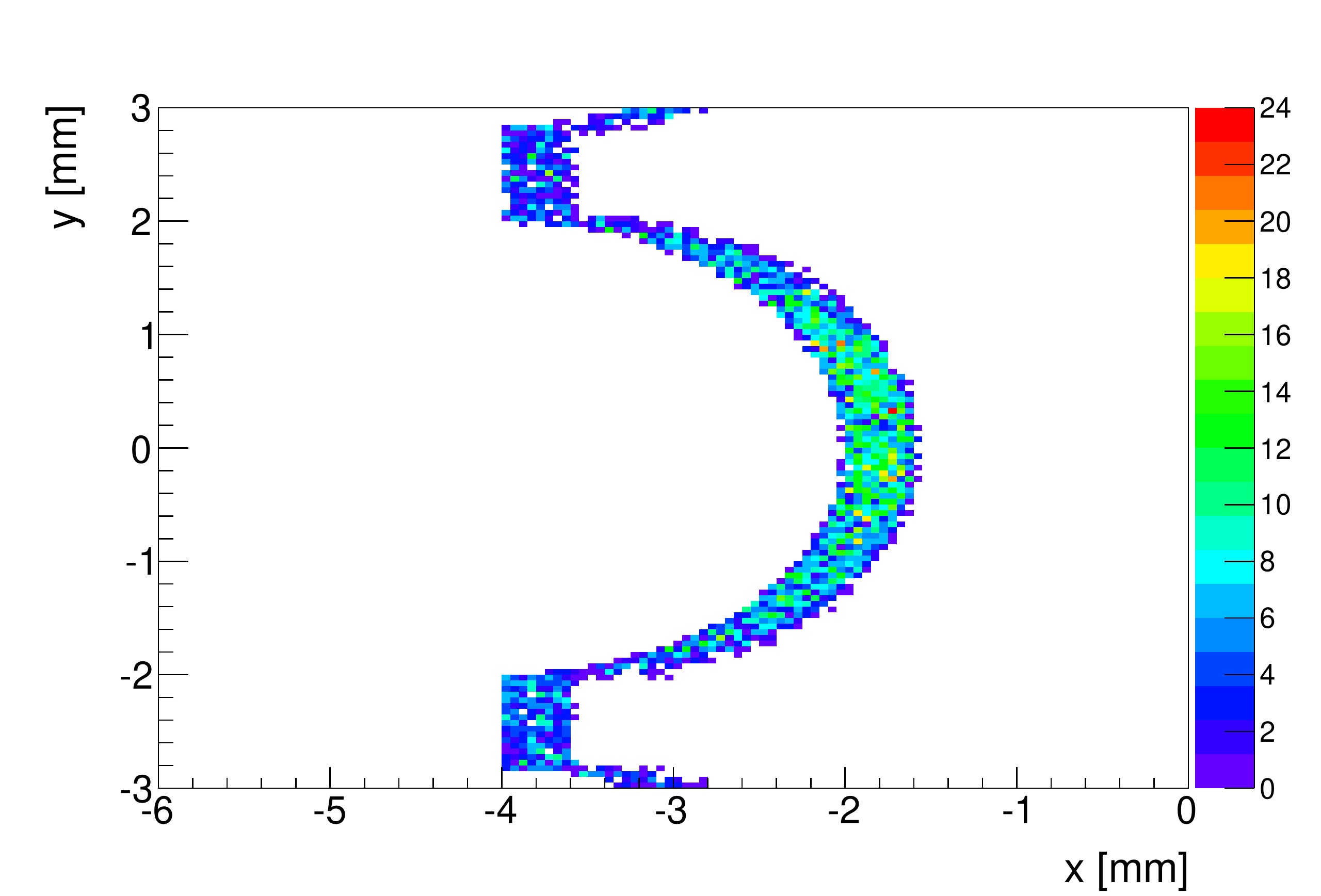}
\caption{Initial positions of accepted muons from the grooved target zoomed in to one groove. Instead of the expected half circular shape the distribution takes on a crescent form thereby reducing the surface volume gain from the grooves.}
\label{fig_InitMuPosFromGroove}      
\end{figure}

The first geometry explored is a radially grooved target where equidistant grooves, placed parallel to the proton beam, are staggered on either side of the target surface, so preserving the overall thickness. The basic idea being to increase the available surface area for surface muon production. No significant improvement over the standard target was observed (see Table~\ref{tab_Enhancement_Factors}). While the grooves increased the geometric surface volume by up to 45\% not all of this volume is useful for small angular acceptance beam lines as the surface becomes too steep for the surface muons with their limited range to still exit the surface volume. This can be seen in Fig.~\ref{fig_InitMuPosFromGroove} which shows the initial positions for accepted surface muons. Instead of the expected half circular shape the distribution takes on a crescent form thereby reducing the surface volume gain from the grooves.

\begin{figure}
\includegraphics[width=1.0\columnwidth]{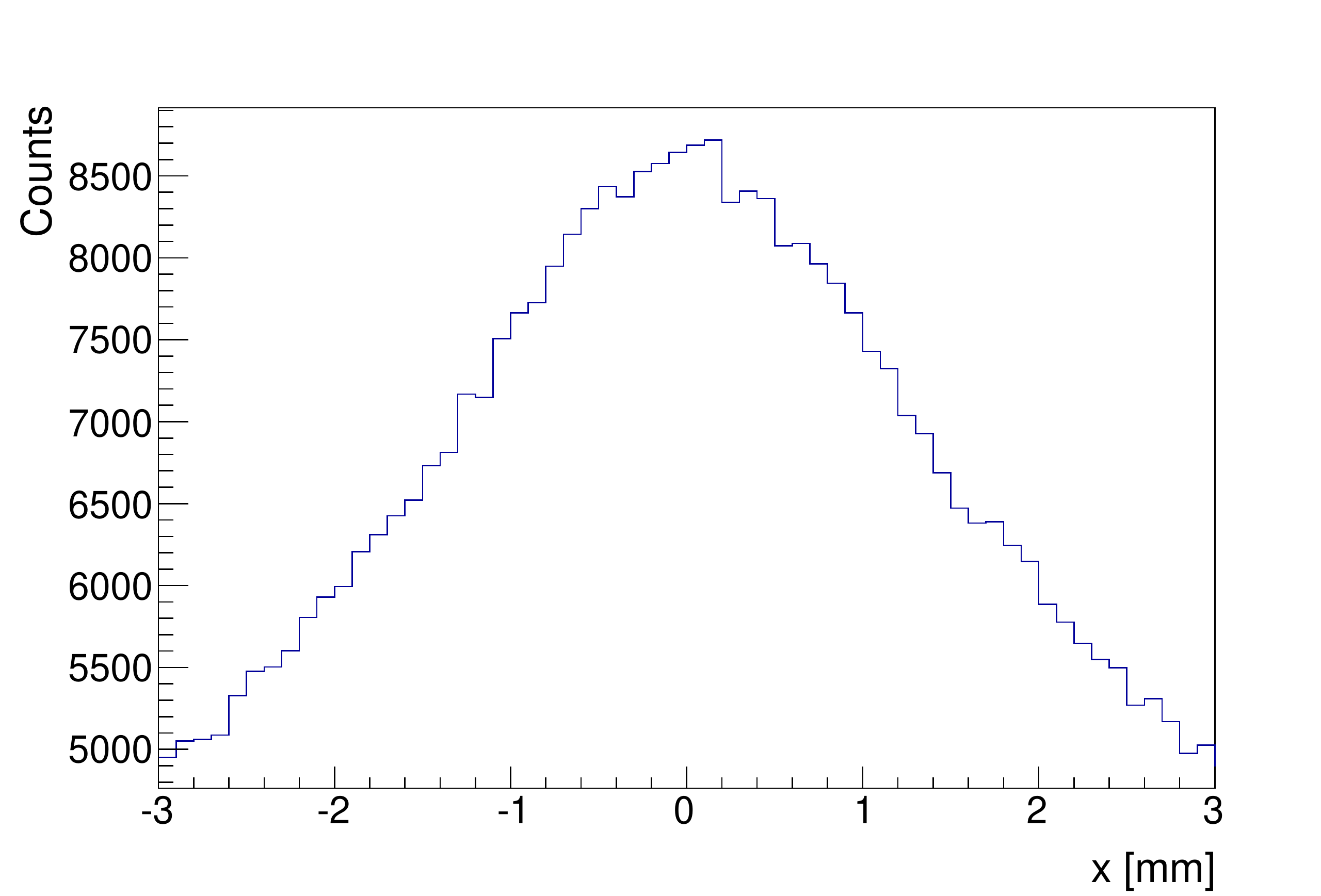}
\caption{Pion stop density through the PSI standard target E in arbitrary units from one side to the other and integrated along its length. While the pion stop density is lowest at the sides where surface muons can actually escape the target it is approximately 70\% higher  in the center.}
\label{fig_PionStopDistX}      
\end{figure}

The small enhancement factors still achieved for the grooved target stem from the fact that the pion stop density is not constant throughout the target. Figure~\ref{fig_PionStopDistX} shows the pion stop density through the target from one side to the other and integrated along its length. While the pion stop density is lowest at the sides where surface muons can actually escape the target it is approximately 70\% higher in the centre. This is due to the fact that the lowest energy pions with only small ranges in the target are stopped very close to the proton path thereby leading to a higher stopping density.

The second geometry investigated is a trapezoidal target with an initial transverse width of 4~mm that increases linearly to 6~mm at the forward end. The basic idea behind this geometry is to exploit the higher pion stop densities close to the centre of the target while still providing the full target length for the bulk of the protons and a somewhat reduced length for the tails of the proton beam. This geometry resulted in a 15\% enhancement to muon rates at $90^{\circ}$ to the target, but a 2\% loss to the backward direction (see Table~\ref{tab_Enhancement_Factors}). The loss in the backward direction is due to the much reduced area of the backward face of the trapezoid target that cannot be recovered by the gains from the side face. The geometry performs even worse for the forward direction for which the surface muon contribution from the side faces is much reduced.

To resolve the inefficiencies of the trapezoidal target and better preserve the proton beam characteristics, a forked target was investigated such that the full proton beam passes through 40~mm of material at every position in the transverse plane. Three sections of target are placed along the proton axis, one upstream centred on the proton beam and two sections placed downstream and offset by the width of the upstream section (see Fig.~\ref{fig_TgEGeometries}). Each section has a width of 2~mm and length of 40~mm, resulting in a total target length of $\sim$~80~mm (depending on overlap). A muon rate enhancement of 45\% in the $90^{\circ}$ direction and a 14\% increase in the backward direction was observed for this geometry. As in the case of the trapezoidal target this geometry has a negative impact on the forward direction.

The final target geometry investigated is a large slab-like target rotated by some angle as can be seen in Fig.~\ref{fig_TgEGeometries}. The length of target material along the proton axis of 40~mm is maintained regardless of rotation angle by scaling the thickness of the slab accordingly. The total length of the slab is independent of the material budget so long as the 40~mm in the beam direction is maintained. We studied the dependence of the enhancement factors as a function of the length of the slab for a rotation angle of $10^{\circ}$. While the length influences the forward and backwards directions only weakly, it is a stronger driver in the sideways extraction. However, after a length of about 75~mm one is already close to saturation (10\% below maximal enhancement for the sideways and $<$5\% below maximal for the backwards and forwards directions). To give the largest enhancement factors achievable we chose to perform our simulations with a length of 150~mm.

\begin{figure}
\includegraphics[width=1.0\columnwidth]{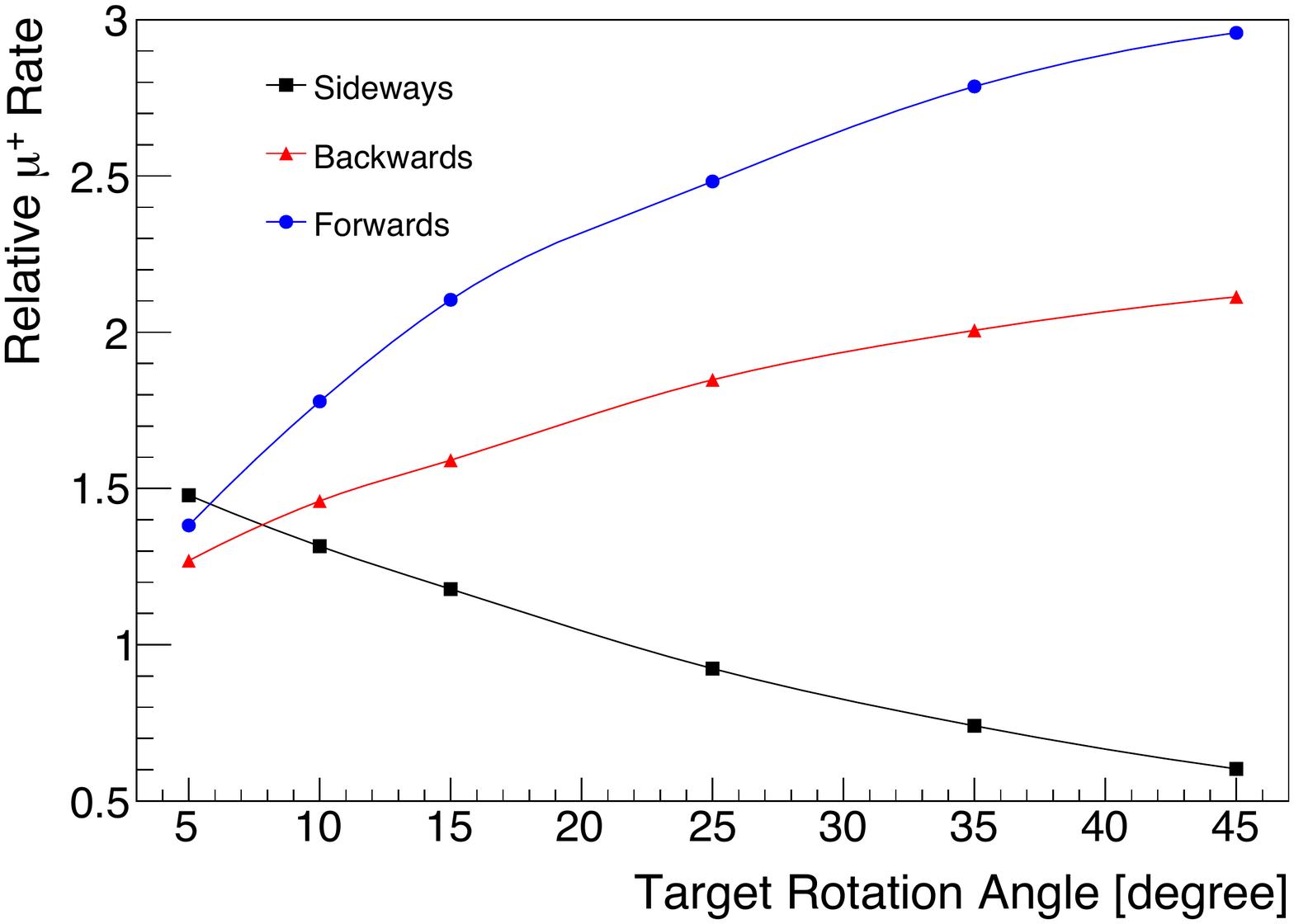}
\caption{Enhancement factors in the three directions studied as a function of the rotation angle of the slab target. The length of the slab is fixed at 150~mm. Rotation angles between 5 and 10 degrees yield roughly equal gains in all three directions.}
\label{fig_SlabMuonRateRotation}      
\end{figure}

The rotation angle strongly influences the enhancement factors that can be achieved in the various beam lines (see Fig.~\ref{fig_SlabMuonRateRotation}). For a target rotation of $10^{\circ}$ a good compromise for the various beam lines can be found, with enhancement factors of 28\% at $90^{\circ}$, 40\% in the backward and even 63\% in the forward direction being observed. The gain in surface muon rates is twofold: i) Depending on the rotation the target can become essentially thinner leading to higher pion stop densities at the surface. Additionally, the area where the protons enter and leave the target and where the pion stop density is highest are distributed across the full surface. ii) As the length of the slab is decoupled from the material budget of the proton beam (40~mm in the case of target E) one can essentially increase the overall length and gain an additional increase in rate.

\begin{figure}
\includegraphics[width=1.0\columnwidth]{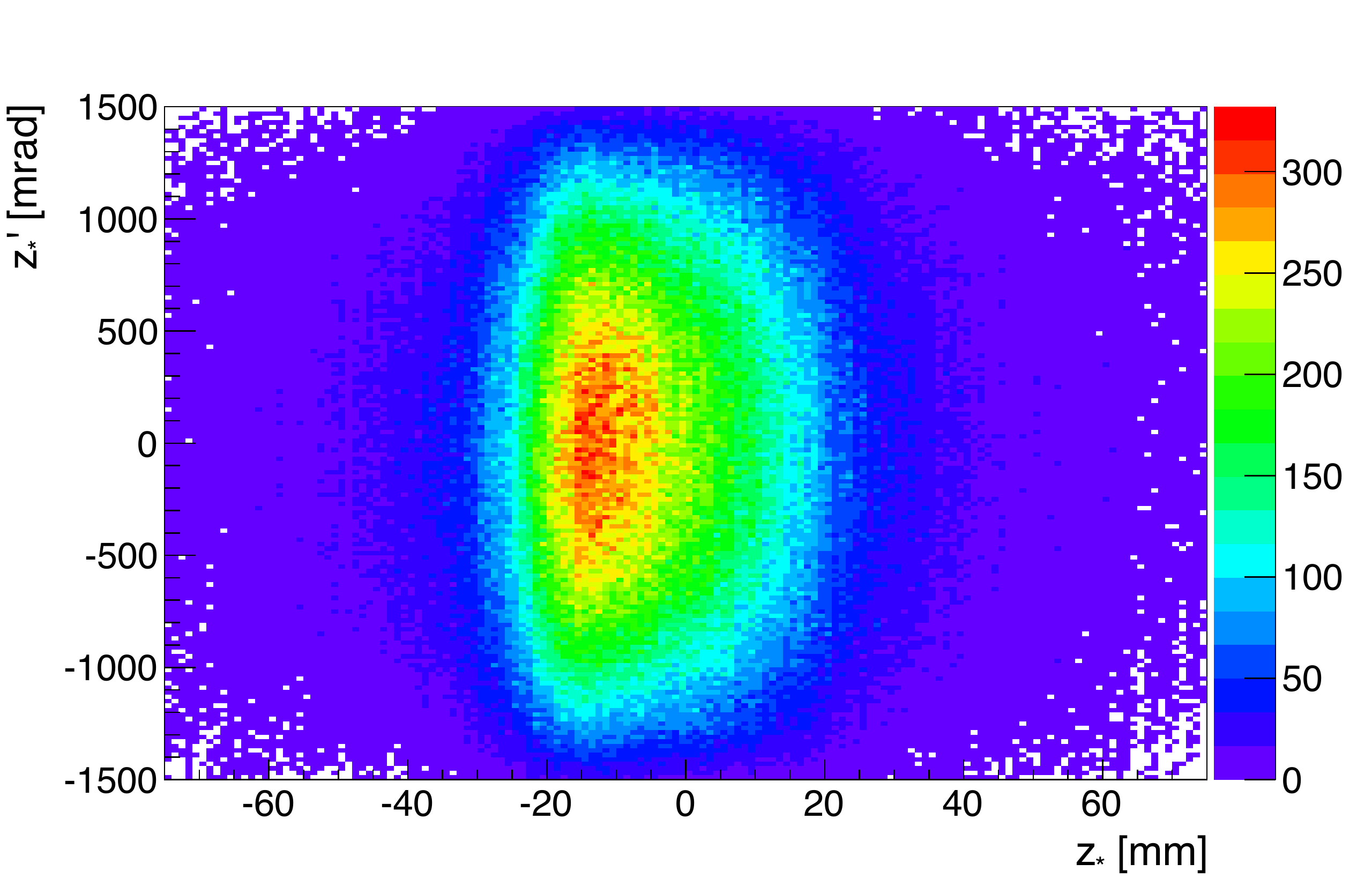}
\caption{Size and divergence distribution along the backward face of the rotated slab target.}
\label{fig_SizeDivergenceRotatedTg}      
\end{figure}

Figure~\ref{fig_SizeDivergenceRotatedTg} shows the size and divergence of emitted surface muons along the backward face of the rotated slab target. The coordinate system is rotated around the $y$-axis such that the rotated $z_*$-axis is parallel to the side of the rotated slab. As in the case of the standard target the divergences are calculated with respect to the $x_*$-direction. The corresponding root-mean-square values are
\begin{eqnarray}
{z_*}_\mathrm{rms} & = & 19.8 \, \mathrm{mm} \nonumber \\
{z'_*}_\mathrm{rms} & = & 673 \, \mathrm{mrad} \nonumber \\
y_\mathrm{rms} & = & 8.3 \, \mathrm{mm} \nonumber \\
y'_\mathrm{rms} & = & 673 \, \mathrm{mrad} 
\end{eqnarray}
with similar values for the forward side. These dimensions are taken along the face of the rotated slab and thus will need to be projected onto the different extraction channels -- sideways, backwards, forwards -- according to their direction and the rotation of the slab.

\begin{figure}
\includegraphics[width=1.0\columnwidth]{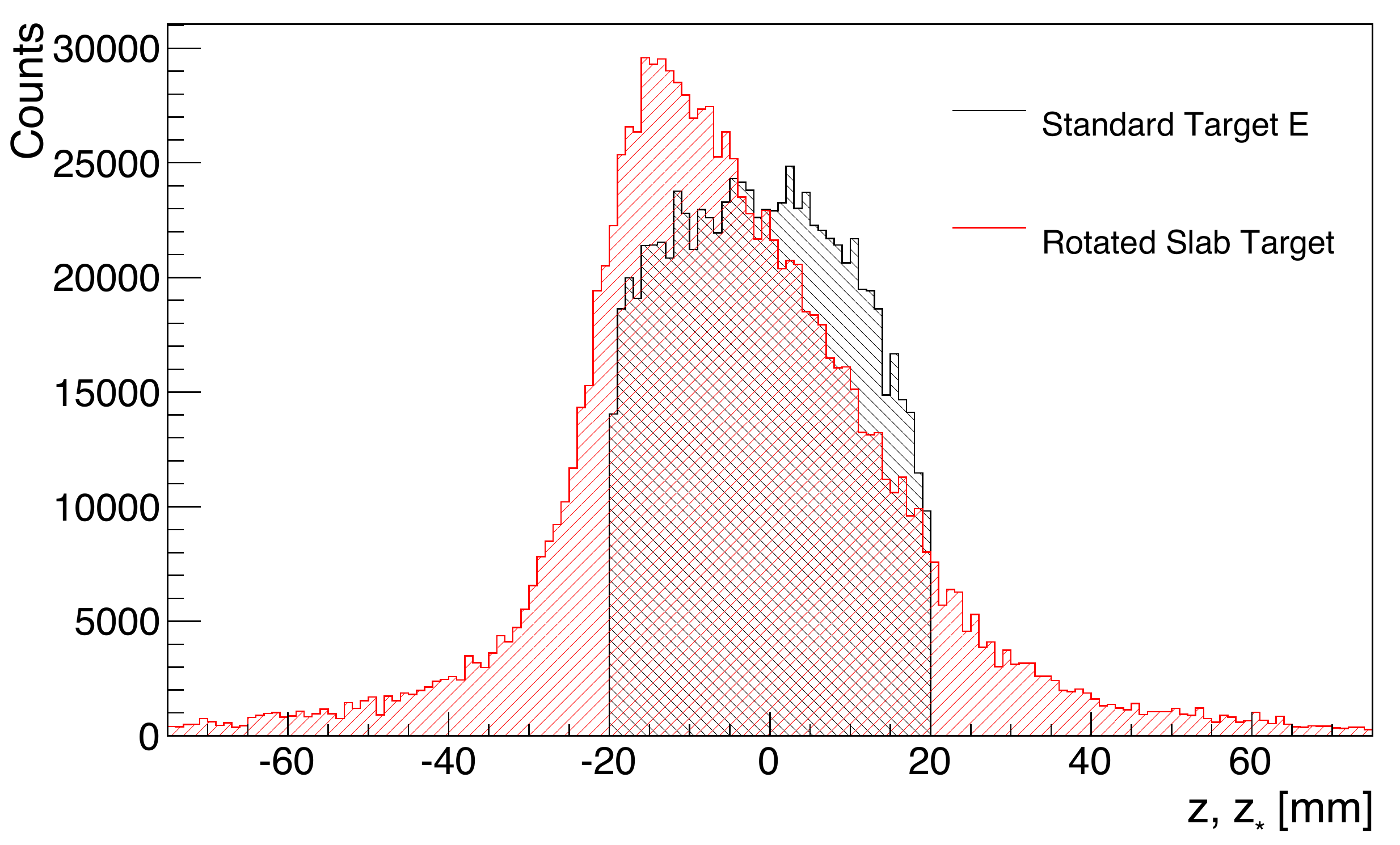}
\caption{Distribution of emitted surface muons along the side face of the standard target and the backward face of the rotated slab target for the same number of protons on target. The overall increase in surface muons for the rotated slab target stems from a somewhat higher peak density and an effectively increased size of the target.}
\label{fig_MuonInitZposition}      
\end{figure}

Figure~\ref{fig_MuonInitZposition} shows a comparison of the position of emitted surface muons along the side face of the standard target and the backward face of the rotated slab target. This comparison corroborates the arguments given above for the overall increase in surface muon rate coming from two factors: i) from a peak in the intensity distribution around the point where the protons enter or leave the target and ii) from an effectively enlarged source compared to the standard target.

\begin{table}
\centering
\begin{tabular}{c c c c}
\hline
\hline
Geometry & Sideways & Backwards & Forwards \\
\hline
grooved & 1.02 & 1.00  & 0.97 \\
trapezoid & 1.15 & 0.98 & 0.79 \\
fork & 1.45 & 1.14 & 0.79 \\
rot. slab & 1.28 & 1.40 & 1.63 \\
\hline
\hline
\end{tabular}
\caption{Enhancement factors for the various geometries (see Fig.~\ref{fig_TgEGeometries}) and directions compared to the standard target E. The rotated slab target yields the best overall enhancement while at the same time being a mechanically simple solution. Statistical errors of the simulation are $\sim$1\% for the sideways and backward directions and $\sim$5\% for the forward direction.}
\label{tab_Enhancement_Factors}
\end{table}

By comparing the various enhancement factors possible for the different geometries, the rotated slab target is clearly superior to the others and offers the additional advantage of being mechanically simpler. The option of a slab target is currently under consideration at PSI.

\section{Optimization of Target Material}\label{sec_MaterialOptimization}

In addition to the geometrical shape of the target its material obviously also plays a major role in the generation of surface muons: Firstly by directly altering the production yield of pions, secondly it has an influence on the number of pions stopped in the target and finally, on the range of muons from stopped pions that can escape from the target.

Effectively, the surface muon rate $I_{\mu^{+}}$ is determined by the pion stop density $\rho_{\pi^+}$, the muon range $r_{\mu^+}$ and the length $l$ of the target for a given material. 
\begin{equation}
I_{\mu^{+}} \propto  \rho_{\pi^+}  r_{\mu^+}  l
\end{equation}
The pion stop density in turn depends on the pion yield of the target material given by its number density $n$ and cross section $\sigma_{\pi^+}$ as well as its stopping power for pions $(\mathrm{d}E/\mathrm{d}x)_{\pi^+}$:
 \begin{equation}
\rho_{\pi^+} \propto n \sigma_{\pi^+}  \left(\frac{\mathrm{d} E}{\mathrm{d} x}\right)_{\pi^{+}}
\end{equation}
Similarly the range of muons is proportional to the inverse of the stopping power:
 \begin{equation}
r_{\mu^+} \propto  \frac{1}{ \left(\frac{\mathrm{d} E}{\mathrm{d} x}\right)_{\mu^{+}} }
\end{equation}
For a surface muon target station that needs to preserve the material budget in the proton beam due to, e.g., a spallation neutron target at its downstream end, requires that the length of the target must be scaled accordingly. Both the energy loss of the protons in the target and the multiple scattering scale linearly with the number of electrons seen by the protons \cite{Oli14}: $n_e = nZl$ with $Z$ being the atomic number. For the inelastic proton-nucleus interactions the scaling is less straightforward but one such parametrization is a scaling approximately proportional to $nA^{2/3}l$ \cite{Tri96} with $A$ being the mass number. For simplicity we continue with the simple scaling based on the number of electrons and normalizing the length of the target to the length $l_\mathrm{C}$ of the standard graphite target with number density $n_\mathrm{C}$  given by
\begin{equation}\label{eq_lengthscale}
l = \frac{ n_\mathrm{C} 6}{n Z} l_\mathrm{C} \, .
\end{equation}

\begin{figure}
\centering
\includegraphics[width=1.0\columnwidth]{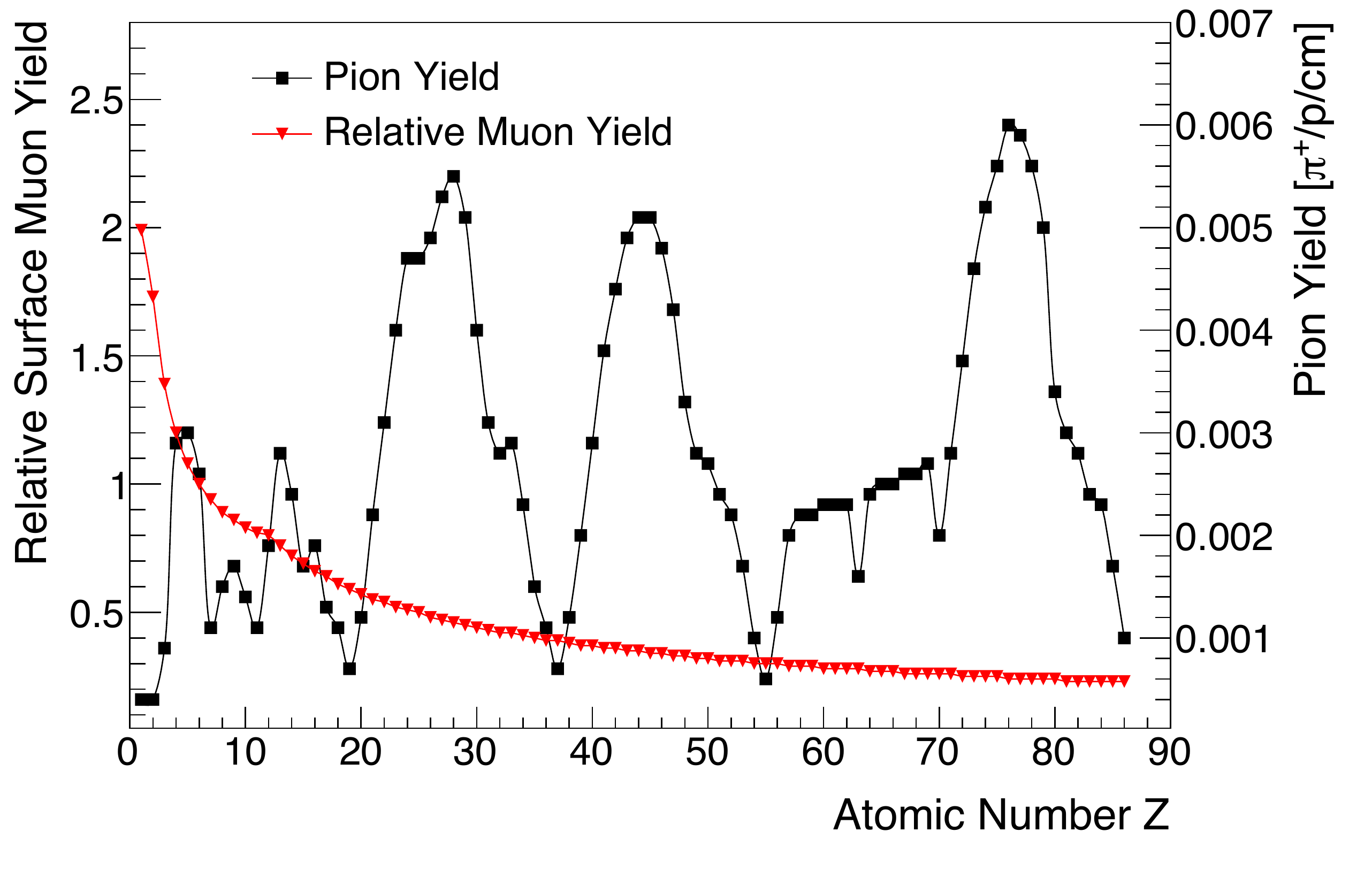}
\caption{The relative surface muon yield and absolute pion yield at a proton energy of 585~MeV as a function of atomic number $Z$. Liquid densities are assumed for elements that are gaseous at normal temperature and pressure.}
\label{fig_RelMuYieldZ}
\end{figure}

Combining the above factors results in a surface muon rate relative to carbon that scales as
\begin{equation}\label{eqn_relmuyield}
I_{\mu^{+}}^\mathrm{rel} \propto n \sigma_{\pi^{+}} \left(\frac{d E}{d x}\right)_{\pi^{+}}  \frac{1}{ \left(\frac{\mathrm{d} E}{\mathrm{d} x}\right)_{\mu^{+}} }  \frac{n_\mathrm{C} 6}{n Z} l_\mathrm{C} \, .
\end{equation}
It is interesting to note the different scalings of the above equation. The pion production cross section $\sigma_{\pi^+}$ approximately scales as $Z^{1/3}$ as can be seen from Eq.~(\ref{eq_totXS}). However, the number density is the dominating factor in the pion yield versus $Z$ as can be seen in Fig.~\ref{fig_RelMuYieldZ}. Due to the fact that the pions and muons have similar masses and stopping power contributions, their effects approximately cancel out in Eq.~(\ref{eqn_relmuyield}). This means that even for a target that does not have to respect the material budget requirements and can thus neglect the length scaling of Eq.~(\ref{eq_lengthscale}) the maximum gain with respect to carbon is only a factor 2 to 2.5 as in the case of, e.g., nickel or tungsten.

The combined $Z$-scaling of Eq.~(\ref{eqn_relmuyield}) is approximately 1/$Z^{2/3}$ and low-$Z$ elements are thus favoured for targets that maintain the proton beam parameters as can be seen in Fig.~\ref{fig_RelMuYieldZ}. While carbon already has a quite low $Z$ value, beryllium would be an option but generally is not suited due to safety reasons such as high evaporation rates and mechanical stress at powerful proton drivers \cite{Hei02}. However, an alternative could be the two carbides boron carbide (B$_{4}$C) and beryllium carbide (Be$_{2}$C). Simulations taking into account the correct length scaling give gains of 10 and 14\%, respectively. As they are ceramics the two materials are temperature resistant and hard but suffer from brittleness. While boron carbide is routinely used in nuclear reactors as control rods \cite{Bar58} there is, to our knowledge, no documented use as a target at a powerful accelerator. At J-PARC there is currently work under way to use silicon carbide as a target material~\cite{Nak15} this should potentially shed light on the performance of such ceramic materials as targets.

\vspace*{1 cm}
\section{Conclusions}
We have investigated the possibilities to improve the surface muon rates at meson factories. With the proton beam power already in the MW-regime attention has to be turned to the optimization of target stations and beam lines. With the combined capture and transport efficiencies of traditional beam lines being of $\mathcal{O}(1\%)$ there is certainly large potential in the optimization of such beam lines. However, this is beyond the scope of this work and we focused here solely on the target side.

By implementing our own pion production cross sections based on a parametrization of existing data into the \textsc{Geant4} simulation package we obtained a reliable tool to predict yields of surface muons at target stations of meson factories. The uncertainty of the simulations are typically in the 10\% range stemming from the inaccuracies of the parametrization and the precision of the measured cross sections.

Firstly, we investigated a novel idea of using a neutron spallation target as a source for surface muons. While the spallation target outperforms standard targets in the backwards direction by more than a factor 7 it is not more efficient than standard targets viewed under 90$^\circ$. Due to the added complications of extracting surface muons from close to a spallation target and the additional disadvantage of having a larger initial phase space, a spallation target is only marginally suited as a source for surface muons.

In the second part of this work we carefully examined potential gains possible over the standard, box-like target employed at PSI from modifications to its geometry and material. It was found that a rotated slab target performs much better with gains of 30 - 60\% possible. An additional gain of 10\% could be achieved from novel target materials such as boron carbide. 

Put into perspective a gain of 50\% would correspond to effectively raising the proton beam power at PSI by 650~kW, equivalent to a beam power of almost 2~MW, without the additional complications such as increased energy and radiation deposition into the target and its surroundings.

\begin{acknowledgments}
Fruitful discussions with Daniela Kiselev, Davide Reggiani, Thomas Rauber, Otmar Morath, Stephane Sanfilippo, Vjeran Vrankovic, Alexander Gabard, Marco Negrazus, Beat Amrein, Thomas Prokscha, and Mike Seidel are gratefully acknowledged. This work was supported under SNF grant 200021\_137738.
\end{acknowledgments}

%

\appendix*
\section{Parametrized Cross Section of Ref.~\cite{Bur89}}
We have modified somewhat the original parametrization found in Ref.~\cite{Bur89} and give here the relevant changes and parameters. Especially we modified the high-energy behavior of the parametrization to not follow a Gaussian shape but to fall off exponentially. In addition, some of the parameters were changed for better agreement between data and the parametrization -- especially for hydrogen, in our case not treated separately.
\begin{eqnarray}\label{eq_XShe}
T_{\pi^+} & \le & \bar{T}(\theta,Z,T_p) + \sigma (\theta,Z,T_p) \,: \nonumber \\
\frac{\mathrm{d}^2\sigma_\mathrm{HE}}{\mathrm{d}\Omega \mathrm{d}T_{\pi^+}} & = & A(\theta,Z,T_p) \exp \left[- \left( \frac{\bar{T}(\theta,Z,T_p) - T_{\pi^+}}{\sqrt{2} \sigma (\theta,Z,T_p)}  \right)^2 \right]   \nonumber \\
& &  \frac{1}{1+ \exp \left[ \frac{T_{\pi^+} - T_F}{B} \right]} \quad \left[ \mu \mathrm{b/(MeV\; sr)} \right] \nonumber \\
& & \nonumber \\
T_{\pi^+} & > & \bar{T}(\theta,Z,T_p) + \sigma (\theta,Z,T_p) \,: \nonumber \\
\frac{\mathrm{d}^2\sigma_\mathrm{HE}}{\mathrm{d}\Omega \mathrm{d}T_{\pi^+}} & = & A(\theta,Z,T_p) e^{-\frac{1}{2}} \nonumber \\
& & \exp \left[- \left( \frac{T_{\pi^+} - (\bar{T}(\theta,Z,T_p) +\sigma (\theta,Z,T_p) )}{n(\theta) \sigma (\theta,Z,T_p)}  \right) \right]   \nonumber \\
& &  \frac{1}{1+ \exp \left[ \frac{T_{\pi^+} - T_F}{B} \right]} \quad \left[ \mu \mathrm{b/(MeV\; sr)} \right]
\end{eqnarray}

The parameters corresponding to Eq.~(\ref{eq_XShe}) are given by the following equations. The energies $T_p$ and $T_{\pi^+}$ are taken in MeV and the angle $\theta$ in degrees.
\begin{eqnarray*}
\bar{T}(\theta,Z,T_p)  & = & 48 + 330 \exp \left( -\frac{\theta}{T_A (Z,T_p)} \right) \\
\sigma (\theta,Z,T_p) & = & \sigma_A (Z ,T_p) \exp \left( -\frac{\theta}{85} \right)  \\
T_A (Z,T_p) & = & \frac{T_A^{730} (Z) (T_p-585) -T_A^{585} (Z) (T_p-730) }{730-585}  \\
T_A^{585} (Z) & = & \left\{ \begin{array}{ll}
28.9 & \textrm{1 $\leq$ Z $<$ 9} \\
26.0 & \textrm{9 $\leq$ Z $<$ 92} 
\end{array} \right. \\
T_A^{730} (Z) & = & \left\{ \begin{array}{ll}
34.2 & \textrm{1 $\leq$ Z $<$ 9} \\
29.9 & \textrm{9 $\leq$ Z $<$ 92} 
\end{array} \right. \\
\sigma_A (Z,T_p) & = & \frac{\sigma_A^{730} (Z) (T_p-585) -\sigma_A^{585} (Z) (T_p-730) }{730-585}  \nonumber\\
\sigma_A^{585} (Z) & = & \left\{ \begin{array}{ll}
130 & \textrm{1 $\leq$ Z $<$ 9} \\
135 & \textrm{9 $\leq$ Z $<$ 92} 
\end{array} \right. \nonumber\\
\sigma_A^{730} (Z) & = & \left\{ \begin{array}{ll}
150 & \textrm{1 $\leq$ Z $<$ 9} \\
166 & \textrm{9 $\leq$ Z $<$ 92} 
\end{array} \right. \nonumber\\
B & = & 50 \nonumber\\
T_F & = & T_p-140-2B \nonumber\\
A(\theta,Z,T_p) & = & N(Z) \sum_{n=1}^5 a_n B_n \nonumber\\
a_1 & = & 27 - 4 \left( \frac{730-T_p}{730-585} \right)^2 \nonumber\\
a_2 & = & 18.2 \nonumber\\
\end{eqnarray*}
\begin{eqnarray}
a_3 & = & 8 \nonumber\\
a_4 & = & 13+(Z-12)/10 \nonumber\\
a_5 & = & 9+(Z-12)/10 - (T_p-685)/20 \nonumber\\
N(Z) & = & c_0 Z^{1/3}+ \sum_{m=1}^3 c_m (\ln Z)^m Z^{1/3} \nonumber\\
c_0 & = & 0.8851 \nonumber\\
c_1 & = & -0.1015 \nonumber\\
c_2 & = & 0.1459 \nonumber\\
c_3 & = & -0.0265 \nonumber\\
n(\theta) & = & 0.4+0.7 \theta /140 
\end{eqnarray}
The B-splines $B_i$ are defined over the range 0 to 180~degrees and follow the knot sequence (0, 0, 0, 30, 70, 180, 180, 180).

While the agreement is generally good for all elements the case of hydrogen is special and agreement between the parametrization and data somewhat worse.

While the total cross section can be obtained by integration of the double differential cross section there exists also, a computationally much simpler approximation \cite{Bur89}, that is used in our simulations and assumes a simple linear behaviour between values close to the pion production threshold and the measurements at 585~MeV and 730~MeV. With the proton kinetic energy in MeV the total cross section is then given in millibarns as:

\begin{widetext}
\begin{eqnarray}\label{eq_totXS}
\sigma (Z,T_p) & = & \left\{ \begin{array}{ll}
\sigma^{585} (Z) (T_p-325)/(585-325) & 325 \le T_p < 585 \\
\sigma^{585} (Z) + ( \sigma^{730} (Z) - \sigma^{585} (Z) )(T_p-585)/(730-585) & 585 \le T_p < 800
\end{array} \right. \nonumber \\
\sigma^{585} (Z) & = & \left\{ \begin{array}{ll}
9.70 & Z=1 \\
28.5 (Z/6)^{1/3}(0.77+0.039Z) & 2 \le Z < 12 \\
19.65 Z^{1/3} & 12 \le Z \\
\end{array} \right. \nonumber \\
\sigma^{730} (Z) & = & \left\{ \begin{array}{ll}
13.50 & Z=1 \\
35.0 (Z/6)^{1/3}(0.77+0.039Z) & 2 \le Z < 12 \\
24.50 Z^{1/3} & 12 \le Z 
\end{array} \right.
\end{eqnarray}
\end{widetext}

\end{document}